# Title

Adaptability of non-genetic diversity in bacterial chemotaxis

# Authors
Nicholas W Frankel, William Pontius, Yann S Dufour, Junjiajia Long, Luis Hernandez-Nunez, Thierry Emonet*

Department of Molecular Cellular and Developmental Biology, Yale University, New Haven, CT
*correspondence: thierry.emonet@yale.edu

# Abstract

Bacterial chemotaxis systems are as diverse as the environments that bacteria inhabit, but how much environmental variation can cells tolerate with a single system? Diversification of a single chemotaxis system could serve as an alternative, or even evolutionary stepping-stone, to switching between multiple systems. We hypothesized that mutations in gene regulation could lead to heritable control of chemotactic diversity. By simulating foraging and colonization of *E. coli* using a single-cell chemotaxis model, we found that different environments selected for different behaviors. The resulting trade-offs show that populations facing diverse environments would ideally diversify behaviors when time for navigation is limited. We show that advantageous diversity can arise from changes in the distribution of protein levels among individuals, which could occur through mutations in gene regulation. We propose experiments to test our prediction that chemotactic diversity in a clonal population could be a selectable trait that enables adaptation to environmental variability.

# Impact statement

Using an experimentally constrained model, we show that *Escherichia coli* faces fitness trade-offs in chemotaxis, and that adaptation of phenotypic diversity through altered gene regulation permits populations to resolve these trade-offs.

# Funding

This research was supported by the James S. McDonnell Foundation (award no. 220020224), the Paul Allen foundation (award no. 11562), the National Institute of Health (grant no. 1R01GM106189), and the Yale University High Performance Computing Center, which is partially funded by National Science Foundation grant #CNS 08-21132.

# Introduction

*Escherichia coli* uses a single chemotaxis protein network to navigate gradients of chemical attractants and repellents, as well as gradients of temperature, oxygen, and pH[1] (Figure 1A). The core of the network is a two-component signal transduction system that carries chemical information gathered by transmembrane receptors to flagellar motors responsible for cell propulsion. A second group of proteins allows the cells to physiologically adapt to changing background signal levels, enabling them to track signal gradients over many orders of magnitude. While different receptors allow cells to sense different signals, all signals are then processed through the same set of cytoplasmic proteins responsible for signal transduction and adaptation. This horizontal integration may impose conflicting demands on the regulation of these core decision-making components because signals can vary in time, space, and identity. In this study, we examine to what extent cell-to-cell variability in abundance of these core proteins may help resolve such conflicts.

The cell uses flagella to explore its environment in a run-and-tumble fashion[2]. Counterclockwise rotation of the flagella promotes the formation of a helical bundle that propels the cell forward in a run. Clockwise rotation tends to disrupt the bundle, interrupting runs with brief direction-changing tumbles. The fraction of time a motor spins clockwise, or clockwise bias, controls the frequency of tumbles and thus plays a central role in chemotactic behavior. Tumble frequency increases monotonically with clockwise bias until the latter reaches about 0.9, at which point cells tumble nearly twice a second and are essentially stationary[3]. It has been observed that clonal cells, grown and observed under the same conditions without stimulation, will differ substantially in clockwise bias[4].

The central logic of *E. coli* chemotaxis is to transiently decrease clockwise bias in response to an increase in attractant signal (Figure 1B). This approach allows cells to climb gradients of attractants by lengthening runs up the gradient (Figure 1C). The adaptation process that maintains receptor sensitivity is mediated by the covalent modification of the chemoreceptors through addition and subtraction of methyl groups by the enzymes CheR and CheB, respectively. Like clockwise bias, the timescale of this adaptation process has been observed to vary among clonal cells[5]. The intracellular levels of these proteins are known to change both adaptation timescale and clockwise bias[6].

Chemoreceptor activity is communicated to the motors via phosphorylation of the response regulator CheY to form CheY-P by the receptor-associated kinase CheA. CheZ opposes the action of CheA by dephosphorylating CheY-P. Consequently, the balance of CheA and CheZ affects clockwise bias. The total amount of CheY in the cell determines the range of possible CheY-P levels, and due to noise in the expression of CheY[7] this dynamic range will likewise vary between clonal cells.

These three phenotypic parameters—clockwise bias, adaptation time, and CheY-P dynamic range—are the main determinants of how *E. coli* performs chemotaxis. These in turn depend on the quantities of chemotaxis proteins within each individual cell. Hence, the copy numbers of these proteins directly determine the ability of the individual to navigate its environment. Since all signals are processed through the same core proteins, this dependency should be independent of the type of signal being followed.

As such, cell-to-cell variability in protein abundance is likely a major contributor to the observed non-genetic behavioral diversity in clonal populations (Figure 1D, 1st and 2nd panels). Various mechanisms can contribute to such variability, including noise in gene expression[8]. Random segregation of proteins during cell division probably plays a role as well[9] and may impose a lower bound on minimum variability attainable[10]. Chemotaxis genes are chromosomally organized in operons—that is, expression of multiple genes are driven by common promoters. This genetic architecture ensures that noise in the activity of shared promoters will affect the expression of multiple genes in a correlated manner, conserving the ratios of proteins from cell to cell despite variations in total amounts[11]. Correlation in protein noise has been experimentally shown to be important in determining chemotactic performance[12]. Combined with the negative integral feedback design of the protein network, this conservation of protein ratios greatly reduces the occurrence of cells with unacceptable parameter values—for instance, those that only run or only tumble—and maintains the precision of the physiological adaptation process[6, 7, 13, 14]. For these and other reasons[15-19], chemotaxis in *E. coli* is often said to be robust.

Within this range of acceptable behaviors, however, substantial variability exists, and the fact that this variability has not been selected against raises the question of whether it might serve an adaptive function. Population diversity is known to be an adaptive strategy for environmental uncertainty[20-22]. In this case of chemotaxis, this would suggest that different cells in the population may hypothetically have behaviors specialized to navigate different environments (Figure 1D, $2^{nd}$ and $3^{rd}$ panels). Indeed, past simulations[23-25] have shown that the speed at which cells climb exponential gradients depends on clockwise bias and adaptation time, and experiments[26] using the capillary assay—an experiment that tests cells' ability to find the mouth of a pipette filled with attractant—have shown that inducing expression of CheR and CheB at different levels changes the chemotactic response. In order to understand the impact of these findings on population diversity, we must place them in an ecological context.

Relatively little is known about the ecology of *E. coli* chemotaxis, but it is probable that they, like other freely swimming bacteria, encounter a wide variety of environments, from gradients whipped up by turbulent eddies[27] to those generated during the consumption of large nutrient caches[28, 29]. In each case, variations in environmental parameters, such as in the amount of turbulence, the diffusivity of the nutrients, or the number of cells, will change the steepness of these gradients over orders of magnitude[27, 30, 31]. Still other challenges include maintaining cell position near a source[32], exploration in the absence of stimuli[33], navigating gradients of multiple compounds[34], navigating toward sites of infection[35], and evading host immune cells[36].

Each of these challenges can be described in terms of characteristic distances and times, for example the length-scale of a nutrient gradient, or the average lifetime of a nutrient source, or the characteristic time- and length-scales of a flow. Chemotactic performance, or the ability of cells to achieve a spatial advantage over time, will depend on how the phenotype of the individual matches the length- and time-scales of the environment. Considering the variety of scales in the aforementioned challenges, and the fact that all must be processed by the same proteins (Figure 1A), it would seem unlikely that a single phenotype would optimally prepare a population for all environments, potentially leading to performance trade-offs (Figure 1D, panel 3) wherein mutual optimization of multiple tasks with a single phenotype is not possible.

Cellular performance will have an impact on fitness (i.e. reproduction or survival) depending on "how much" nutrient or positional advantage is required to divide or avoid death. Therefore, selection that acts on chemotactic performance could transform performance trade-offs into fitness trade-offs (Figure 1D, panels 3 and 4), which are known to have direct consequences for the evolution of diversity[20-22, 37]. Selection that favors top performers disproportionately to intermediate performers could hypothetically transform a weak performance trade-off into a strong fitness trade-off.

Fitness trade-offs can lead to the development of multiple biological modules[38]. Some modules, like new limbs, may be permanent fixtures, while others, like metabolic pathways, may be switched on and off, either in response to the environment as it changes, or stochastically in anticipation of environmental fluctuation. This latter case, often called "bet-hedging[39]," is a strategy used by bacteria to avoid extinction from antibiotic stress during infection[40] and has evolved in the laboratory under fluctuating selection[41]. In these examples, environmental extremes lead to discrete partitioning of the population. Is there an intermediate case, a possible evolutionary stepping-stone, in which a single function is continuously diversified in the population without the formation of a wholly different state?

In this study, we seek to determine to what extent advantageous diversity can be created from a single biological network, as well as the possible mechanisms that may permit adaptation of such diversity in response to selective pressures. Due to the lack of quantitative information about the details of the natural environments of *E. coli* that would be relevant to chemotaxis, our goal is not the exact reconstruction of the distribution of challenges experienced by *E. coli* in the wild, but rather to use bacterial chemotaxis as a system to study the interactions between population diversity and environmental trade-offs. Although different signals are sensed by different receptors, the cell interprets all signals using the same set of proteins. Since we are interested in the relationship between cellular dynamics and the length- and time-scales of the environment, this allows us to simplify our study by focusing on different gradients of a single attractant type. Our findings should apply to different signal identities as well.

For this study, we must be able to translate an individual cell's protein concentrations into its fitness in different environments (Figure 1D). Chemotaxis in *E. coli* is a system uniquely well-suited to this purpose. The wealth of molecular and cellular data that has been gathered by different research groups over the last several decades makes it one of the best-characterized systems in biology for the study of single-cell signal transduction and behavior. A key result of this past research is a molecular model of *E. coli* chemotaxis, which accounts for the interactions of all of the proteins in the network, that we will fit simultaneously to many experimental data sets. From this model, we will be able to calculate phenotypic parameters such as adaptation and clockwise bias as a function of protein concentrations (Fig 1D, map from $1^{st}$ to $2^{nd}$ panel).

We will then simulate the performance of virtual *E. coli* cells with these phenotypic parameters to characterize any trade-offs that *E. coli* faces in performing fundamental chemotactic tasks (Fig 1D, map from $2^{nd}$ to $3^{rd}$ panel). These tasks will be parameterized by characteristic lengths (distance to source) and times (time allotted). A wide range of these environmental parameters will be explored to ensure that a full spectrum of cell–environment interactions are investigated.

We will measure the performance of cells in the environments and apply different ecological models of selection to assign fitness. In doing so, we will examine how performance trade-offs give rise to fitness trade-offs (Fig 1D, map from $3^{rd}$ to $4^{th}$ panel). Finally, we will use a model of population diversity based on noisy gene expression to determine whether changing genetic regulation could allow populations to achieve a collective fitness advantage.

**Results**

*A mathematical model maps protein abundance to phenotypic parameters to behavior*

The first step in creating a single-cell conversion from protein levels into fitness was to build a model of the chemotaxis network. We began with a standard molecular model of signal transduction based explicitly on biochemical interactions of network proteins. We simultaneously fit the model to multiple datasets measured in clonal wildtype cells by multiple labs[4, 7, 42]. Along with previous measurements reported in the literature, this fitting procedure fixed the values of all biochemical parameters (i.e. reaction rates and binding constants), leaving protein concentrations as the only quantities determining cell behavior (Methods, Supplementary File 1 – Parameter Values).

The fit took advantage of newer single-cell data not used in previous models that characterize the distribution of clockwise bias and adaptation time in a clonal population[4]. In order to fit this data, we coupled the molecular model with a model of variability in protein abundance, adapted from Lovdok et al.[11] (Methods). In this model, the abundance of each protein is lognormal-distributed and depends on a few parameters that determine the mean abundance and the extrinsic (correlated) and intrinsic (uncorrelated) noise in protein abundance (details of the model discussed further below)[8]. By combining these components, our model simultaneously fit the mean behavior of the population[7, 42] and the noisy distribution of single-cell behaviors[4] (Figure 2 – figure supplement 1). In all cases, a single set of fixed biochemical parameters was used, the only driver of behavioral differences between cells being differences in protein abundance.

Given an individual with a particular set of protein levels, we then needed to be able to calculate the phenotypic parameters: adaptation time, clockwise bias, and CheY-P dynamic range. To do so we solved for the steady state of the model and its linear response to small deviations in stimuli relative to background (Methods). This produced formulae for the phenotypic parameters in terms of protein concentrations.

For simplicity, we did not model the interactions of multiple flagella. Rather, we assumed that switching from counterclockwise to clockwise would initiate a tumble after a lag of 0.2 s that was required to account for the finite duration of switching conformation. A similar delay was imposed on switches from tumbles to runs. In this paper we only consider clockwise bias values below ~0.9, because above this value cells can spend many seconds in the clockwise state[3]. During such long intervals, non-canonical swimming in the clockwise state can occur. In this case, the chemotactic response is inverted and cells tend to drift away from attractants[43]. This behavior is therefore maladaptive for the cell; however, it is only observed in mutant cells[3, 43]. In experiments with wildtype cells, this regime is not observed[4] because of the robust architecture of the network[7] (Methods).

Using the definitions for adaptation time, clockwise bias, and CheY-P dynamic range, we reduced the molecular model into a phenotypic model written in terms of phenotypic parameters rather than protein levels (Methods). Simulating the step-response of the molecular model with a given set of protein levels matched the behavior of the phenotypic model with corresponding phenotypic parameters (Figure 2 – figure supplement 2). Because there are half as many phenotypic parameters as different proteins, the phenotypic model made it computationally possible to explore large ranges of behavior in the simulations we describe in the next section.

*Different environments require different behaviors*

To characterize chemotactic trade-offs faced by *E. coli*, we began by investigating which chemotactic phenotypes performed best in different ecological tasks. Here, we defined a phenotype as a particular set of values of the phenotypic parameters: adaptation time, clockwise bias, and CheY-P dynamic range. We used the phenotypic model to simulate the behavior of individual phenotypes in various environments and measured the performance of each phenotype based on metrics appropriate to each ecological challenge. In total, these steps provided us with a direct mapping from individual protein levels to chemotactic performance in the ecological tasks we describe below (Figure 1D).

*E. coli*, like other commensals and pathogens, must survive relatively nutrient-poor environments outside the host until it can colonize a new host. An important ecological parameter in this situation is the characteristic distance between resources, which sets the typical signal length-scale a bacterium must navigate. When a source is close, the challenge might be to climb steep gradients and to stay near the source. In contrast, when the environment is sparse, the ability to explore and navigate shallow gradients may be more important. Another important ecological parameter is the characteristic time-scale of changes in the environment, which dictates the allotted time a bacterium has to navigate its environment. As this time becomes shorter (e.g. when bacteria must take advantage of nutrient patches in moving flows[27, 44]) chemotactic performance becomes more important. For these reasons, we parameterized environments in terms of distances and times. The range of values was chosen such that at one extreme, cells begin at the source, and at the other, the distance and time requirements are so stringent that reaching the source is only possible by swimming randomly (i.e. pure diffusion).

We considered two tasks. The first is a foraging challenge in which a spherical parcel of nutrient appears at a certain distance from the cell and immediately begins to diffuse away. This occurs, for instance, upon lysis of a unicellular eukaryote[28]. The location of the parcel is unpredictable and could be close or far. Each cell in the simulation accumulated nutrient by collecting an amount proportional to the concentration at its position at every time-step. Performance was defined as the amount of nutrient acquired (Figure 2A) within a certain time limit. For simplicity we assumed that consumption by an individual is small enough not to have an impact on the gradient itself. Feedback of populations onto the shape of the gradient certainly plays a role in many ecological scenarios and could be considered in this framework in the future.

The second environment recapitulates a colonization task, in which a colonization site opens up at a random distance from the cell and immediately starts releasing an attractant signal by diffusion. This case is analogous to the classic capillary experiment[45] and may have relevance to infection by species such as uropathogenic *E. coli*[46]. We approximated the site as a persistent spherical zone with a non-depleting concentration of attractant. Performance was defined by minimizing the time to reach the site, equivalent to maximizing the reciprocal of the arrival time, before a global time limit, which may be determined ecologically by the carrying capacity of the site or the periodic purging of the area around the site (Figure 2D). Cells unable to reach the colonization site by that time were given an infinite arrival time and consequently a zero performance value. Later, we consider the reduction of this time limit as an ecological factor.

For each ecological task, we scanned different environmental parameters (the distance at which the source appears, the time allotted, and the source concentration [Figure 2 – figure supplements 3, 5, and 7, respectively]) and simulated the performance of different phenotypes. For each phenotype and environment, 6,000–30,000 replicate trajectories were averaged together to quantify performance as a function of phenotype and environment. We began with no constraints or correlations between phenotypic parameters and scanned them independently; later we consider the effect of biological constraints on phenotypic distributions.

When a nearby source appeared, cells in the foraging challenge immediately experienced high nutrient levels and were challenged to maintain their position despite having been exposed to a large increase in signal. Successful cells had high clockwise bias, which curtails long runs, and short adaptation time, which mitigates large responses (Figure 2B). These cells in a way are defeating chemotaxis and motility both: to stay rooted, they tumble constantly and have a fast adaptation time that reduces the duration of response. If chemotactic populations are preparing for unexpected types of environments but have uniformly turned on expression of chemotaxis and motility genes, cells with these phenotypes could potentially function as if those processes were turned off, without having to introduce a genetic on-off switch. Conversely, when a source appeared farther away, cells had to use longer runs to reach the expanding front of the gradient and long adaptation times to integrate the weaker signals at its tails (Figure 2C). If time is further limited, this far-source effect is exaggerated (Figure 2 – figure supplement 5).

The case of colonization was similar, except that shorter adaptation times were favored overall as compared to foraging. This was because the gradient geometry is much steeper in the vicinity of the source due to its persistently high concentration, and climbing that final part of the gradient was required for colonization (Figure 2, D inset versus A inset). Climbing steep gradients requires fast adaptation to stay abreast of quickly changing background levels.

The source concentration played a minor role in colonization; however, when foraging less concentrated sources, the favored strategy for far distances inverts from low to high clockwise bias, indicating that at that point little can be gained from motility—in fact higher motility may move the cell away from the source (Figure 2 – figure supplement 7). The dynamic range of CheY-P has a negligible effect on cell performance so long as it is sufficiently high as to ensure that the response of CheY-P to kinase activity is linear and does not saturate (Figure 2 – figure supplement 8). For this reason, when we discuss optimal performance in the subsequent analysis, we assume that the total amount of CheY molecules in the cell has been selected to be high enough to avoid these limitations (Methods).

In both challenges, the distance at which the source appeared substantially changed which phenotypes outperformed the others. In general, distant sources required lower clockwise bias and longer adaptation time than nearby ones (Figure 2, C compared to B and F compared to E). This becomes even more apparent if we plot the optimal phenotype as a function of source distance (Figure 2 – figure supplement 4). These results are consistent with our recent study[25] that used an analytical model to predict the velocity of cells climbing static one-dimensional gradients and detailed the mechanistic basis of performance differences between phenotypes. There, we demonstrated a trade-off wherein steep gradients required fast adaptation time and high clockwise bias for optimal velocity, whereas shallow gradients required slow adaptation time and low clockwise bias. Our present simulations of ecological tasks show that this trade-off also exists in more complex chemotactic scenarios. The dependence of the optimal phenotype on the environment follows the same trend in the previous analytical model as it does in our current simulation results, wherein simulations of distant sources are similar to simple shallow gradients and nearer sources are analogous to steeper gradients.

*Trade-off strength and population strategy depend on the nature of selection*

Using two ecological tasks, we have shown that a single phenotype cannot perform optimally in all environmental conditions. To understand the consequences of these trade-offs, we must analyze whether they are weak or strong. Such analysis will reveal in which cases populations should adopt homogenous or diversified strategies, respectively, for optimal collective performance.

For a two-environment trade-off, the fitness of all possible phenotypes in both environments occupies a region in two-dimensional fitness space called the fitness set[47] (Figure 3, gray regions). Specialists in this set will be located at the region's maxima in each axis (red and blue circles). Between the specialists, the outer boundary of the set is called the Pareto front[37]: a group of phenotypes that have jointly optimized both tasks (black line). A generalist phenotype will occupy a position on this front (gray circle). When this front is convex (middle panel), the generalist has higher joint performance. A concave front (right panel), however, is optimized by a mixed strategy of specialists, due to the fact that a combination of specialists (dashed line) will exceed the fitness of any phenotype in the fitness set[20].

Assuming cells have negligible ability to control or predict at what distance the next source will appear, cells are mutually tasked with survival in both near and far sources. As such, we examined trade-offs between pairs of near and far environments to test to what extent cells can cope with environmental variability. In each environment, performance is evaluated on a scale relative to the richness of that environment. That is to say, nearby sources will naturally result in higher performance values than distant ones. Such differences in scale between different tasks do not change the significance of the curvature of the Pareto front; in fact, axes can even have different units and the meaning of the curvature will be the same[37].

Trade-offs in performance arose when cells were required to mutually optimize foraging or colonization of nearby and far away sources (Figure 4). This is a consequence of the fact that unique specialists, defined by different clockwise bias and adaptation time, are needed for each environment (Figure 2 – figure supplement 4). Since these optimal phenotypes are not changed by CheY-P dynamic range as long as it sufficiently high (Figure 2 – figure supplement 8), this phenotypic parameter does not contribute to performance trade-offs. As the disparity between these source distances becomes greater, the front of the trade-off transitions from convex to concave (Figure 4, from A to C for foraging and from D to F for colonization), demonstrating that performance trade-offs in fundamental tasks can be strong when environmental variability is high. Trade-offs become much stronger when the environment turns over rapidly (Figure 2 – figure supplement 6).

Nutrition and arrival time, however, are not themselves equivalent to fitness. Fitness quantifies how these performance metrics would contribute to cellular survival and reproduction. Taking a neutral performance trade-off case for each task type (Figure 4, B and E), we asked the questions: how are performance trade-offs translated into fitness trade-offs, and how does the nature of selection influence their strength?

In the case of foraging, survival depends on the ability to scavenge sufficient nutrition. The metabolic reactions that mediate this survival are nonlinear biochemical processes. Many such reactions follow sigmoidal relationships, like the Hill equation, rather than linear ones. We created a simple metabolic relationship in which the survival probability of an individual cell was expressed as a Hill function with two parameters: the amount of food required for survival, and how strongly survival probability depended on that amount (Figure 5A). To obtain the fitness of a phenotype, we calculated the expected value of its survival by averaging the survival probability of all replicate cells with that phenotype (Methods).

When the nutrition requirement was low and the dependency was weak, the previously neutral trade-off became a weak fitness trade-off (Figure 5B). Increasing the nutrition requirement and dependency imposed stricter selection, which penalized all but the top performers. This transformed the underlying neutral performance trade-off into a strong fitness trade-off (Figure 5C). Therefore, the selection parameters themselves can determine the strength of fitness trade-offs. Discrete transitions between survival outcomes gave qualitatively similar results (Figure 5 – figure supplement 1A–C).

In the case of colonization, individual success was binary: either the colonization site was successfully reached, securing that cell's survival for the near future, or the cell was purged—e.g. consumed by a neutrophil—and left no progeny. We approximate the fitness reward for each replicate as a function of the arrival time that steps down from one to zero after a time limit (Figure 5D).

As in the foraging case, we calculated the colonization fitness, or survival expectation, of a phenotype as the average of the zero and one outcomes of all the individual replicate cells of that phenotype. When the time limit was high, the previously neutral performance trade-off gave rise to a weak fitness trade-off (Figure 5E). When the transition point was lower, the same performance trade-off became a strong fitness trade-off (Figure 5F). These effects mirror those demonstrated in the foraging case. As an alternative calculation of fitness we also considered a continuous reward function, which qualitatively produced the same results (Figure 5 – figure supplement 1D–F).

The common thread between these cases is that the nonlinear relationship between performance and fitness can increase or reduce the fitness difference between the high-performing specialists and intermediate-performing generalist, strengthening or weakening the trade-off, respectively. Thus, whether diversification is advantageous depends not only on performance trade-offs, but also on the selection process, which has the potential to reverse the strength of trade-offs. Understanding fitness trade-offs therefore requires consideration of both performance and selection.

*Genetic control of non-genetic diversity enables populations to resolve trade-offs*

We have identified conditions in which diversified populations have a fitness advantage over homogeneous ones: those in which the environment is highly variable and those in which selection truncates populations to the top performers. While we cannot know with certainty what trade-offs wildtype *E. coli* have experienced, we do know that they exhibit substantial phenotypic heterogeneity in their swimming behavior. As mentioned earlier, our model of bacterial chemotaxis, when combined with a model of cell-to-cell variability in protein abundance, reproduces the variability in adaptation time and clockwise bias measured in a wildtype population (Figure 2 – figure supplement 1). While there are certainly variations in other quantities, such as cell size or number of flagella, the fit of our model suggests that noise in protein levels is a plausible driver for behavioral diversity in *E. coli* chemotaxis.

Since phenotypic selection can alter variability in protein abundance[48], we asked the question of whether selection on genetic regulatory features of the chemotaxis network could serve as an adaptive mechanism capable of shaping diversity in protein abundance, and thus phenotypes, to resolve trade-offs. Such features include the organization of the genes on the chromosome and the sequences of ribosomal binding sites (RBSs) and promoter regions. Selection for individuals with mutations in these features would give rise to adaptation of the distribution without changing highly-conserved network proteins. In our model of gene expression, such alterations were realized through changes in the levels of extrinsic and intrinsic noise and the mean expression level of each protein. We first varied these parameters individually to investigate their effects on phenotypic diversity (Figure 6).

Intrinsic noise results in diversification of protein ratios (Figure 6A). Intrinsic noise can be reduced when multiple genes are expressed from one operon—as are the core chemotaxis genes *cheRBYZ*[11]. Intrinsic noise is increased when translation of a protein is highly stochastic or when individual proteins are driven by different promoters that are decoupled. When we compared populations that had low or high intrinsic noise (Figure 6B, light blue and dark blue, respectively) we observed that high intrinsic noise resulted in many cells having clockwise bias near 0 or 1 and therefore being non-chemotactic (Figure 6C, dark blue). Reducing intrinsic noise resulted in more cells having phenotypic parameters within the functional range, consistent with previous experimental findings[7, 11, 12]. We also observed an inverse correlation between clockwise bias and adaptation time that is known to arise from the architecture of the network[4, 49] (Figure 6C, light blue).

Altering the strength of an RBS changes the mean protein ratios, resulting in a shift in the mean phenotype of the population without directly affecting population variability (Figure 6D). Experimentally, mutations in RBSs at the single-nucleotide level are known to have profound effects on expression levels of chemotaxis genes in *E. coli*[15]. We illustrated this by increasing the mean level of CheR (Figure 6E, pink). CheR is responsible for receptor methylation, so increasing its mean level decreased the mean adaptation time (Figure 6F). There was also an increase in mean clockwise bias due to the fact that increasing CheR relative to CheB increases the steady-state methylation level (in spite of the mitigating effect of the CheB-P feedback), leading to higher clockwise bias.

Extrinsic noise (Figure 6G) arises both from variations in global factors in the cell, such as differences in the number of ribosomes or errors in protein partitioning during cell division, as well as from the noisiness of promoters that drive multicistronic operons[51-53]. Reduction of extrinsic noise, which for example could occur through stronger feedback control on a promoter[10], resulted in a population with a tighter, more homogenous distribution of phenotypes (Figure 6H–I, red). Hence, through pathway-specific mutations in the promoter or its regulators, we predict that clonal populations could approach a more generalist-like distribution or a more multi-specialist-like distribution. However, noise cannot be eliminated entirely[10], suggesting that there may be a fundamental limit to the efficacy of a generalist strategy through the reduction of protein noise.

To determine whether changing these regulatory parameters alone can generate Pareto-optimal population distributions, we numerically optimized population fitness (Methods), allowing only the two noise magnitudes and the mean expression levels to vary. Populations were comprised of individual cells, each having a fitness in each environment. Following previous studies[22], the fitness of the population in a given environment was defined as the average fitness of all of its individuals in that environment. For simplicity we assumed that the population encountered environments one at a time and survived all environments. Therefore the population fitness over all environments was the geometric mean of the population fitness in each environment, weighted by the probability of encountering each environment (Methods). The environments considered were the same as in Figure 5, which include examples of both strong and weak trade-offs for each ecological task.

We used the wildtype level of intrinsic noise obtained in our fit to experimental data (Figure 2 – figure supplement 1) as a lower bound in the optimization. Multiple experimental studies show that wildtype cells reduce intrinsic noise for improved chemotactic function[7, 11, 12], so we inferred that they may be operating near a fundamental lower limit. We also set a lower bound on the total noise level based on experimental measurements in *E. coli* of protein abundance in individual cells over a large range of proteins[54](Methods). This bound is primarily from irreducible extrinsic noise arising from various mechanisms such as the unavoidability unequal partitioning of proteins during cell division. We set an upper bound on mean protein levels to 5 fold above the wildtype mean in order to be within a range of experimentally established observations[7, 50] (Methods).

When we optimized populations for weak trade-off in either foraging or colonization tasks, the resulting populations in both tasks exhibited lower levels of protein noise (Figure 7A for foraging and 7E for colonization, blue points) and lower phenotypic variability (Figure 7B and F), in comparison to populations optimized for the respective strong foraging or colonization trade-offs (Figure 7ABEF, red compared to blue points). In all cases, the spread of individuals in the optimal populations was constrained to the Pareto front (Figure 7CDGH). The spread was more condensed in the weak trade-offs than in the strong trade-off in the same task (Figure 7C compared to D for foraging and G compared to H for colonization)

In the weak trade-off cases, condensation into a single point on the Pareto front was impeded by lower bounds on noise. Even though a pure generalist strategy was unattainable, adjustments in the means and correlations between protein abundance enabled the system to shape the "residual" noise to distribute cells along the Pareto front. This could be a general phenomenon in biological systems: given that molecular noise is irreducible, the best solution is to constrain diversity to the Pareto front. Our results suggest this may be achievable via mutations in the regulatory elements of a pathway.

In the strong foraging trade-off, the optimized population took advantage of the fact that correlated noise in protein levels leads to an inverse relationship between clockwise bias and adaptation time (Figure 7A and B, red) due to the architecture of the network. By capitalizing on this feature, the population contained specialists for near sources, which had higher clockwise bias and shorter adaptation times, and those for far sources, which had lower clockwise bias and longer adaptation time. Cells with clockwise bias above 0.25 were avoided because steep gradients were very short-lived in this challenge.

The strong colonization trade-off also required high clockwise bias for near sources and low clockwise bias for far sources. However, since the gradient in the site's vicinity did not flatten, a short adaptation time was always necessary to climb the final part of the gradient and clockwise bias above 0.25 could become advantageous when the source is close. In order to achieve greater diversity in clockwise bias while keeping adaptation time low, the optimized population for this trade-off had increased intrinsic noise, which diversified protein ratios and disrupted the inverse correlation between clockwise bias and adaptation time (Figure 7E and F, red).

In all cases, selection on regulatory elements of the network resulted in phenotypic diversity being remarkably constrained to the Pareto front. Furthermore, the levels of diversity in these populations are consistent with the sign of the curvature of the Pareto front they occupy. The adaptability of these distributions predicts that genetic alterations to basic regulatory mechanisms may allow clonal cells to resolve multi-objective problems at the population level using a single signaling network. This mechanism could allow populations to cope with the need to navigate diverse environments, or follow diverse signals, without partitioning into discrete subpopulations through the use of switches or modules.

*Potential future experiments suggested by the theory*

Our results could be tested using several types of chemotactic performance experiments. The radial symmetry of our environments makes it possible to use the capillary[26] and plug assays[55], which present cells with a concentrated source of attractant. The soft agar swarm plate assay[12] could be used as well if modified to introduce nutrient solution to one spot of the plate instead of the whole. Microfluidic chemotaxis assays[34] could be constructed using soft lithography to reproduce these environments with a higher level of precision. In each of these cases, the distance between cells and the source and the duration for which the source is presented could be varied, as well as the source concentration. Cells with high performance should be selected, analyzed for their phenotypes and protein abundance, and re-grown, either under continual presentation of the same condition or switching between two or more conditions.

Using these types of experiments, our theoretical results predict several specific outcomes. First, seeding the same clonal population in different assays that have different length- or time-scales should select for different optimal subpopulations with different phenotypic parameters and different levels of protein expression. Such measurements would make it experimentally possible to verify the chemotactic trade-offs we predict. Experimental work using the capillary assay already supports this claim[26].

In the case of laboratory evolution with one selection condition, we predict an eventual shift toward genotypes that suppress population noise, as well as toward mutations in chemotaxis protein RBSs that allow the mean clockwise bias and adaptation time to specialize for this task. In this case, we predict that populations will reduce phenotypic diversity but run into a lower limit of protein noise. These outcomes could be measured by performing single cell phenotype analyses and by re-sequencing the operon. Conversely, alternating selection in different assays or different length- and time-scales may lead to enhanced phenotypic noise and still other RBS mutations. In these cases, whole genome re-sequencing may show alterations to the operon structure or to the master regulators of chemotaxis.

Strains that are evolved in the lab could be compared to the wildtype ancestor in order to gain insight into the types of environments the latter evolved in. Furthermore, investigating phenotypic diversity in wild strains in comparison to domesticated and evolved laboratory strains may uncover differences that reflect the level of environmental diversity faced in their respective lifestyles.

**Discussion**

The chemotaxis system exhibits significant plasticity in the shape of phenotypic distributions, which can provide fitness advantages in chemotactic trade-offs. Such trade-offs arise from environmental variability because the performance of a chemotactic phenotype is sensitive to the length- and time-scales of the environment it must navigate. This dependency is especially strong when time for navigation is limited. Though at this stage we cannot know what distribution of chemotactic challenges wildtype *E. coli* have faced, we do expect trade-offs to arise from the diversity of time- and length-scales in environmental encounters.

Our simulations environments were simplified. They omitted many real-world factors for future studies, such as competition between multiple species, turbulence, and viscosity in environments such as soil or animal mucosa. As new data on these interactions emerge, the framework we introduced could be used to investigate trade-offs and resulting phenotypic distributions. Additionally, interactions with more than two environments are likely to occur and could be analyzed in the same way. Such cases will likely impose more constraints on navigation, giving rise to stronger trade-off problems. Increasing the number of phenotypic parameters would not necessarily alleviate these constraints, which would instead be primarily governed by the distance between the optimal phenotypes for these tasks in phenotypic parameter space.

In our framework, we expanded the traditional genotype–phenotype relationship to consider protein levels separately. While genotype could be broadly defined to include both coding sequences and regulators of noise, separate treatment of protein levels permitted analysis of copy number variability apart from changes in the proteins themselves. This approach could be applied to other signal transduction systems, since variability in the levels of signaling proteins may change behavior as much as changing protein biochemistry.

In this study, we tuned the distribution of protein levels using numerical parameters, but such changes would in fact occur through mutations. Mean expression levels could change via gene duplication, RBS point mutations, mRNA structures, or altered activity of upstream regulators. Phages and recombination events can reorganize genes, changing intrinsic noise relative to extrinsic noise by altering expression correlation. Regulators of promoters can incur mutations that result in negative feedback repression to reduce promoter noise. Protein localization affects partitioning noise, which is interesting since some chemotaxis proteins assemble into discrete membrane-bound clusters while others do not.

In the future, it would be interesting to study the extent to which higher expression levels will result in fitness costs, possibly introducing trade-offs. For instance, physiological adaptation via the enzymatic actions of CheR and CheB consumes cellular resources, imposing metabolic costs that depend inversely on the adaptation timescale[56]. Different media and growth phases alter the expression levels of these proteins[50, 57] and will naturally change the distribution of phenotypes as well—this could be a mechanism for separating protein levels required for chemotaxis from those better suited for growth. In this study, challenges and regrowth occurred in discrete sequential steps and there was no direct inheritance of phenotype. The relative importance of these features will depend on the relationship between their time-scales and those of the environmental challenges[21]. If the time-scale of environmental change is much slower than the time-scale of adaptation, for example, populations will adapt to their current environment rather than the statistics of environmental fluctuations.

A new feature of our conceptual framework is the distinction between performance and fitness. Organisms exhibit many behaviors that, to researchers, are not directly connected to survival and reproduction. These gaps in our understanding inhibit our ability to understand the evolutionary significance of many organismal behaviors. Here, we demonstrated methods for broaching these questions quantitatively, and in so doing uncovered the relevant finding that nonlinearities in selection can strengthen or weaken trade-offs. This will be of general interest to those studying fitness trade-offs since the nature of selection can change the optimality of pure versus mixed population strategies.

While we have used *E. coli* as a model system due to the wealth of experimental data, the framework developed here could be used to extend these questions to other human commensals and pathogens, with the hope of better understanding their ecology and pathogenesis. The closely related chemotaxis system in *Salmonella enterica* is required for virulence[58], as is the substantially different motility system of *Borrelia burgdorferi*[59]. On the other hand, pathogens such as *Pseudomonas aeruginosa* have multiple motility systems to tackle different environments during infection[60]. Phenotypic diversification within a single system may bridge the gap between one system and many by allowing populations to adapt to greater environmental variation without developing a new biological module. Multicellular organisms also exhibit different motion strategies in their constituent cells, from the singular approach of human sperm to the different motility patterns of neutrophils as they navigate the body to sites of infection and capture invading organisms[61]. Our framework could be used to investigate several open questions in such systems: How does behavioral diversity of single cells affect the fitness of the organism, and when is the diversification of a single cell type supplanted by the commitment of a new developmental cell lineage?

From the simplest two-component systems to the most elaborate signal transduction cascades, proteins responsible for sensing environmental signals are usually distinct from those involved in making behavioral decisions. Often, the output of many types of receptor proteins are fed into a much smaller number of signal transduction pathways. While cells can control their sensitivity to different signals by regulating the expression of different receptors, the integration of multiple signals through a central group of proteins will place conflicting demands on those core proteins. Thus, while horizontal integration is beautifully economical and a ubiquitous feature of biological pathways, our study illustrates that it is also likely to introduce trade-offs by design.

Biology is replete with noise. Although the concept of non-genetic individuality may have been initially coined in reference to *E. coli* chemotaxis[5], we now know that many other biological systems exhibit substantial non-genetic cell-to-cell variability, including stem cell differentiation[62], bacterial sporulation[63], and cancer cell response to chemotherapy[64]. Different systems may have different mechanistic drivers that create, constrain, and adapt this variability. In all cases, however, it is conceivable that through genetic changes to drivers of non-genetic diversity, populations of cells may achieve higher collective success in tackling biological trade-off problems. This form of diversity may constitute an evolutionary stepping stone on the path from one to multiple biological modules.

**Methods**

*Single-cell model of chemotaxis under control of a population-level model of gene expression*

Model outline

We created a single-cell model of *E. coli* chemotaxis that models the switching of flagellar motors (Eqs. 1–3), the activity of chemoreceptors (Eqs. 4, 5), and the biochemical actions of the signal transduction and adaptation enzymes (Eqs. 6–10). In order to calculate the phenotypic parameters in terms of protein concentrations, we performed a linear response perturbation analysis (Eqs. 11–16). Taken together, this model (Eqs. 1–16) allowed us to convert concentrations of chemotaxis proteins into time-dependent behavior and phenotypic parameters.

In order to generate different cells with different levels of chemotaxis proteins we used a model of population variability (Eqs. 17, 18). This allowed us to fit the model to multiple data sets measured in wildtype RP437 strain cells (Figure 2 – figure supplement 1). Before performing simulations, we simplified the model by rewriting it in terms phenotypic parameters directly rather than protein concentrations (Eq. 19, 20).

Flagellar motors

Bacterial flagellar motors switch between counterclockwise rotation, associated with relatively straight swimming, and clockwise rotation, associated with periods of tumbling. We model the bacterial flagellar motor as a bistable stochastically switching system[17, 65]. The free energies of the states, and consequently the switching rates between states, are modulated by the concentration of phosphorylated messenger protein CheY, $Y_p$. We assume that the free energy difference between the CCW and CW states is linear in the occupancy of the motor protein FliM by CheY-P. The rates $k_+$ and $k_-$ of switching out of the CW and CCW states, respectively, are then given by

$$k_\pm = \omega_0 \cdot e^{\pm \frac{g}{2} \left( \frac{1}{2} - \frac{Y_p}{Y_p + K_d} \right)}, \quad (1)$$

in which $\omega_0$ sets the maximum rate of motor switching, $g$ sets the scale of the free energy difference, and $K_d$ is the FliM-CheY-P dissociation constant. Instantaneous CW bias $CW$ as a function of CheY-P input is given by

$$CW = \frac{k_-}{k_+ + k_-}, \quad (2)$$

which describes a sigmoidal curve[66]. Here $g$ determines the steepness of the relationship, and $K_d$ sets the location of the midpoint. The noise in the $Y_p$ signal is modeled using a normal distribution $N(Y_p)$ with mean $Y_{p,0}$ and variance, $\sigma_{Y_p}^2$, the time-averaged CW bias $CW_0$ is obtained by averaging the instantaneous CW bias according to

$$CW_0 = \int_{-\infty}^{\infty} CW(Y_p) N(Y_p) dY_p. \quad (3)$$

When the system is either unstimulated or fully adapted to a constant background, the system is said to be at steady state. In such conditions, $Y_p = Y_{p,SS}$, $CW_0 = CW_{SS}$. The "clockwise bias" we refer to in the main text is $CW_{SS}$ and is set by $Y_{p,SS}$ through Eq. (3), along with $\sigma_{Y_p}^2$, which is calculated below under "Linearization of the chemotaxis pathway model." We show how $Y_{p,SS}$ is calculated below under "Molecular model of the chemotaxis pathway."

In our model, we make the simplifying assumption that a switch from counterclockwise to clockwise rotation initiates a tumble (following a 0.2 delays to account for conformation changes) and therefore clockwise bias is approximately equivalent to the tumble bias[17]. Experiments carried with mutants, however, show that, when clockwise bias is above about 0.9, the motors spend enough time in the clockwise state that the flagella adopt right-handed helices which can propel the cell forward in a "clockwise run." Consequently cells with extremely high clockwise bias will swim down gradients of attractants because they will perform a clockwise "tumble" when going up and a counterclockwise "run" when going down[43]. The reason that this switch in behavior happens at extremely high clockwise bias and not symmetrically at clockwise bias 0.5 is not fully understood, but data shows that the residence times in the clockwise state are much shorter than those in the counterclockwise state throughout most of the range of clockwise bias. As such, the data suggests that clockwise state residence times only become long enough for "clockwise running" when CheY-P is at extremely high concentrations and clockwise bias is essentially 1[3]. Consideration of cells with CW bias above 0.9 is not relevant for our study because neither in measurements of wildtype cells [4](see Figure 2 – figure supplement 1 for our fit to that data) nor in our optimized populations (Figure 7) do we see more than 1% of cells with clockwise bias above 0.5. In experiments, cells with clockwise bias above 0.9 are only observed in mutants (e.g. CheB mutant)[3]. In wildtype, negative feedback of the kinase CheA on CheB-P implies that even when CheY or CheR are highly over expressed the level of CheY-P is maintained low enough to avoid that deleterious regime of inverted chemotaxis. As a result we can rule out this effect for the current study.

Bacterial chemoreceptors

Bacteria sense changes in their external environment using transmembrane chemoreceptors. These receptors are sensitive to changes in the concentrations of various chemical stimuli as well as temperature, oxygen levels, and acidity. Receptors respond to stimuli by modulating their rates of switching between active and inactive conformations. Here we model the receptor response to the chemoattractant methyl-aspartate using a Monod-Wyman-Changeux model of mixed complexes of Tar and Tsr receptor types[16, 67]. Each MWC complex consists of $N_{Tsr}$ = 4 Tar and $N_{Tar}$ = 2 Tsr homodimers[16]. Receptors within each complex are assumed to switch in an all-or-none fashion. The free energy of the active conformation is taken to decrease linearly with the methylation level $m_c$ of the complex, as determined experimentally[42]. For this model, the mean activity $a$ of the complex as a function of $m_c$ and the external methyl-aspartate stimulus $L$ is

$$a(m_c, L) = \left(1 + e^{\varepsilon_0 + \varepsilon_1 m_c} f(L)\right)^{-1}, \qquad (4)$$

in which $\varepsilon_0$ and $\varepsilon_1$ are constants, and the function

$$f(L) = \left(\frac{1 + L/K_{Tar}^{off}}{1 + L/K_{Tar}^{on}}\right)^{N_{Tar}} \left(\frac{1 + L/K_{Tsr}^{off}}{1 + L/K_{Tsr}^{on}}\right)^{N_{Tsr}}. \qquad (5)$$

The constants $K_{Tar}^{off}$, $K_{Tar}^{on}$, $K_{Tsr}^{off}$, and $K_{Tsr}^{on}$ characterize the binding of methyl-aspartate to Tar and Tsr in active and inactive conformations. In the models of the chemotaxis pathway below we use $m$ to denote the mean methylation level of all MWC complexes in the cell and take $a(m)$ to be the mean activity of all complexes in the cell, following previous studies[17, 42, 67, 68]. This approximation is equivalent to assuming that the distribution of $m_c$ across the cell is sharply peaked around $m$ or, alternatively, that $a(m_c)$ is linear in $m_c$.

Molecular model of chemotaxis

Receptor activity adapts to persistent stimulus through methylation and demethylation of the receptors by the enzymes CheR and CheB, respectively. In modeling the kinetics of receptor modification, we follow previous work that successfully describes the adaptive response measured in populations in bacteria[42, 49]. In this model, CheR binds preferentially to inactive receptors and CheB to active receptors. The average methylation level $m$ of all MWC complexes therefore evolves according to

$$\frac{dm}{dt} = \frac{2N}{T_{Tot}} \left( \frac{k_r R_{Tot} T}{K_r + T} - \frac{k_b B_{p,Tot} T^*}{K_b + T^*} \right) + \eta_m(t), \tag{6}$$

in which $T_{Tot}$, $R_{Tot}$, and $B_{p,Tot}$ are the total concentrations of receptors, CheR, and phosphorylated CheB in the cell, $K_r$ and $K_b$ are Michaelis-Menten constants characterizing the enzyme-receptor binding, and $k_r$ and $k_b$ are the catalytic rates for receptor methylation and demethylation. $T^*$ and $T$ denote the concentrations of free active and inactive receptors, respectively. Since the number of enzyme-receptor complexes is small relative to the number of receptors, we make the approximations $T^* + T \sim T_{Tot}$ and for the mean activity of the system, $a \sim T^*/T_{Tot}$. We define $N = N_{Tar} + N_{Tsr}$ as the size of the MWC complexes, so $T_{Tot}/2N$ is the total concentration of MWC complexes in the cell. The term $\eta_m(t)$ is a white noise source that introduces spontaneous fluctuations in methylation level. While models of the form of Eq. (6) correctly describe the adaptation dynamics of averaged populations, they generally fail to predict sufficiently high levels of noise[49]. Therefore, we set the intensity of the noise source $\eta_m(t)$ to agree with experimental measurements, as discussed in the next section. Differentiating $a(m, L)$ and using $\partial a / \partial m = \varepsilon_1 a(1-a)$, we may recast Eq. (6) to describe the evolution of the mean activity $a$ of the system:

$$\frac{da}{dt} = \varepsilon_1 a(1-a) \frac{2N}{T_{Tot}} \left( \frac{k_r R_{Tot}(1-a)}{K_r/T_{Tot} + 1 - a} - \frac{k_b B_{Tot} a}{K_b/T_{Tot} + a} \right) + \frac{\partial a}{\partial L} \dot{L} + \varepsilon_1 a(1-a) \eta_m(t) \tag{7}$$

We note that the stimulus term depends on the time derivative of the ligand concentration $\dot{L}$. At steady state the steady state activity $a_0$ is given by $\frac{k_r R_{Tot}(1-a_0)}{K_r/T_{Tot} + 1 - a_0} = \frac{k_b B_{P,Tot} a_0}{K_b/T_{Tot} + a_0}$.

In their active form, receptors promote the autophosphorylation of an associated histidine kinase CheA, which in turn phosphorylates CheB and the messenger protein CheY that regulates the activity of the flagellar motors. The concentration $A_p$ of phosphorylated CheA is then described by

$$\frac{dA_p}{dt} = a_p \left( A_{Tot} - A_p \right) a - a_b A_p \left( B_{Tot} - B_{p,Tot} \right) - a_y A_p \left( Y_{Tot} - Y_p \right), \tag{8}$$

in which $a_p$, $a_b$, and $a_y$ are rate constants, $A_{Tot}$, $B_{Tot}$, and $Y_{Tot}$ are the total concentrations of CheA, CheB, and CheY, and $B_p$ and $Y_p$ are the concentrations of free CheB-P and CheY-P. CheB-P (either free or bound to a receptor) autodephosphorylates at a rate $d_b$ and CheY-P is dephosphorylated by CheZ with a rate $d_z$. The levels of phosphorylated CheB and CheY then follow:

$$\frac{dB_{p,Tot}}{dt} = a_b A_p \left( B_{Tot} - B_{p,Tot} \right) - d_b B_{p,Tot}, \tag{9}$$

and

$$\frac{dY_p}{dt} = a_y A_p \left( Y_{tot} - Y_p \right) - d_z Z_{Tot} Y_p, \tag{10}$$

in which $Z_{Tot}$ is the total concentration of CheZ molecules.

The molecular model depends on the biochemical parameters, $k_r$, $k_b$, $K_r$, $K_b$, $a_p$, $a_y$, $d_z$, $d_b$, $a_b$, which are the same for all cells since we consider isogenic populations (Supplementary File 1 – Parameter Values and Methods section "Constant biochemical parameters of the model" below), and on the molecular abundance of $A_{tot}$, $T_{tot}$, $R_{tot}$, $B_{tot}$, $Y_{tot}$, and $Z_{tot}$.

Linearization of the molecular model

Eqs. (6-10) constitute a nonlinear system $dX/dt = F(X) + S + H$ describing the evolution of $X = (a, A_p, B_p, Y_p)$ in the presence of a stimulus $S$ and noise source $H$. $F$ is a vector-valued function specified by Eqs. (6-10). In the absence of stimulus $S = 0$ the steady state of the system $X_0$ is the solution of $F(X_0) = 0$. For small stimuli and noise levels that induce only small changes $\delta X = X - X_0$ about the steady-state, we may linearize the system to obtain

$$\delta \dot{X} = J \delta X + S + H, \tag{11}$$

in which $J$ is the Jacobian of $F$ evaluated at the steady state $X_0$:

$$J = \begin{pmatrix} \dfrac{\partial \dot{a}}{\partial a} & \cdots & \dfrac{\partial \dot{a}}{\partial Y_p} \\ \vdots & \ddots & \vdots \\ \dfrac{\partial \dot{Y}_p}{\partial a} & \cdots & \dfrac{\partial \dot{Y}_p}{\partial Y_p} \end{pmatrix}_{X=X_0}. \tag{12}$$

The eigenvalues of $J$ for the model Eqs. (6-10) are generally negative, indicating that the system relaxes to its steady state after small perturbations. The methylation reactions of Eq. (6) are slow relative to the phosphorylation reactions described by Eqs. (8-10) and therefore effectively determine the rate of this relaxation. This rate is given by the largest (least negative) of the eigenvalues $\lambda$ of $J$, which we use to define the relaxation time scale of the system

$$\tau = -\frac{1}{\max(\lambda)}. \tag{13}$$

We note that this rate sets the rate of relaxation to both external stimuli and intrinsic noise[4, 68].

Magnitude of spontaneous fluctuations

Measurements[4] have indicated that the variance $\sigma_{Y_p}^2$ of intrinsic temporal fluctuations in CheY-P scales linearly with the relaxation time scale $\tau$, according to

$$\sigma_{Y_p}^2 = C\tau, \tag{14}$$

with $C = 3.89 \times 10^{-3}$ μM²/s. We assume these fluctuations arise solely from fluctuations in the mean methylation level $m$. Therefore, for a value of $\tau$ calculated from the reaction constants and protein concentrations in a given cell, we choose the intensity of the noise source $\eta_m(t)$ in Eq. (6) so that $\sigma^2_{Y_p}$ and $\tau$ satisfy Eq. (14). Specifically, we first calculate $\tau$ for a given cell and calculate the corresponding variance $\sigma^2_{Y_p}$ from Eq. (14). Since the phosphorylation processes in Eqs. (8-10) are fast relative to the methylation process of Eq. (6), they may be considered to be in the steady-state and Eq. (6) is effectively a one-dimensional Ornstein–Uhlenbeck process. We therefore can relate $\sigma^2_{Y_p}$ to the variance of the intrinsic temporal fluctuations in the methylation level $\sigma^2_m$ by

$$\sigma_m = \sigma_{Y_p} \left( \frac{dY_p}{da} \right)^{-1}. \tag{15}$$

Here, $dY_p/da$ is calculated from the function $Y_p(a)$, Eq. (16) below, obtained from solving Eqs. (6-10) at steady state, as described fully in the next section. Since $\tau$ corresponds to the relaxation time of the methylation process in Eq. (6), we then use $\tau$ and $\sigma^2_m$ to set the intensity of the noise source $\eta_m(t)$ according to

$$\langle \eta_m(t)\eta_m(t') \rangle = 2\sigma^2_m / \tau \delta(t-t') \tag{16}$$

in which $\delta(t)$ is the Dirac delta.

Gene expression model

The reaction rates are assumed to be the same for all cells since the population we consider is isogenic. The total numbers of protein, however, do change from cell to cell and their distribution over the population are determined using a stochastic gene expression model described in this section.

We adapted a model[11] of noisy gene expression that produces individual cells each with an individual numbers of proteins $\mathbf{P} = [A_{Tot}\ W_{Tot}\ R_{Tot}\ B_{Tot}\ Y_{Tot}\ Z_{Tot}\ T_{Tot}]$,:

$$\mathbf{P} = \xi_{ex}\mathbf{P_0} + \sqrt{\xi_{ex}}\mathbf{A} \cdot \text{diag}(\eta\mathbf{P_0}) \cdot \boldsymbol{\xi_{in}}, \tag{17}$$

where $\mathbf{P_0}$ is the corresponding vector of mean protein levels in the population, $\boldsymbol{\xi_{in}}$ and $\xi_{ex}$ are the intrinsic and extrinsic noise generators[8], respectively, $\eta$ is the scaling of the intrinsic noise (taken to be a constant for all proteins for simplicity), and $\mathbf{A}$ is the translational coupling matrix[11], a lower triangular matrix of correlation coefficients $a_{ij}$ between proteins $i$ and $j$. The intrinsic noise $\boldsymbol{\xi_{in}}$ is a vector of normally-distributed random variables with mean zero and variance one, providing individual uncorrelated noise sources for each protein. The extrinsic noise $\xi_{ex}$ is a single lognormal-distributed random variable that provides correlated noise to all proteins together given by

$$\xi_{ex} = \frac{1}{e^{\frac{1}{2}(\omega\ln(10))^2}} e^{\xi\omega\ln(10)}, \tag{18}$$

where $\xi$ is a normally-distributed with mean zero and variance one, and $\omega$ is a scaling parameter for the extrinsic noise.

Since many proteins of the pathway assemble into ultrastable membrane-associated complexes[69, 70], the individual protein levels generated from the noisy gene expression model was further constrained by taking into account the experimentally observed stoichiometry: CheW docks to Tar and Tsr with 2:12 stoichiometry, CheA docks to receptor-associated CheW with 2:2 stoichiometry, and CheA is synthesized in two isoforms, CheA$_L$ and CheA$_S$, with a 45:22 ratio[50]; only the CheA$_L$ form has kinase activity, CheZ docks to CheA$_S$ and has more activity than un-docked CheZ[71-74], so we assume for simplicity that only docked CheZ has significant activity. These relationships were used to determine the number of functional receptor complexes on a per-cell basis, producing final effective levels of $Z_{tot}$, $A_{tot}$, and $T_{tot}$ to be used in the single cell model described above. The extra copies of proteins not in complexes did not participate in the signaling.

Constant biochemical parameters of the model

Before conducting any simulations or analysis, we performed a one-time fitting routine to fix the biochemical parameters ($k_r$, $k_b$, $K_r$, $K_b$, $a_p$, $a_y$, $d_z$, $d_b$, $a_b$), which we assume are the same for all cells since we consider isogenic populations. Most of these parameter values were fixed from previous experiments (SI Table 1) *except* for $k_r$, $k_b$, $K_r$, $K_b$ and $a_p$, which we fit to data. To perform the fitting, we set the population mean protein levels **P**$_0$ to the wildtype levels[50] except where noted below to match the experiment. Since the intrinsic and extrinsic noise scaling parameters $\eta$ and $\omega$ are unknown for wildtype cells, those were allowed to change along with the biochemical parameters that were being fitted. In summary, the biochemical parameters $k_r$, $k_b$, $K_r$, $K_b$, and $a_p$, and the gene expression parameters $\eta$ and $\omega$ were used as the 7 fit parameters. After the fit was performed, $\eta$ and $\omega$ were allowed to vary again (i.e. in Figure 6 and the population optimization for Figure 7), but the biochemical parameters ($k_r$, $k_b$, $K_r$, $K_b$, $a_p$, $a_y$, $d_z$, $d_b$, $a_b$) were fixed permanently for all populations in all contexts.

As fit-data we used (i) measurements of the histogram of CW bias in a wildtype population and the adaptation times associated with each bin[4] (Figure 2 – figure supplement 1A); (ii) measurements of the population-average CW bias as a function of fold overexpression of the mean protein levels **P**$_0$[7] (Figure 2 – figure supplement 1B); and (iii) population-averaged relationship between receptor activity level and methylation rate [42] (Figure 2 – figure supplement 1C). To fit the later data we used our molecular model to simulate and reproduce the time-dependent experimental method used in these experiments: we simulated the response of CheY-P levels within populations of 100 immobilized cells to exponential ramps of ligand. We used the same ramp rates and "strain" ($\Delta tsr$ cells) as in the experiments.

We used a cost function that was simply the sum squared error of all data points and corresponding model/simulation results. We minimized the cost function, allowing the 7 fit-parameters mentioned above to vary, using MATLAB's pattern search optimization algorithm.

Importantly, with a single set of parameter values, the resulting model agrees well with multiple experimental measurements of both single cells and populations of cells from several laboratories (Figure 2 – figure supplement 1). Compared to single-cell measurements of the histogram of CW bias in the population, the model produces a similar spread (Figure 2 – figure supplement 1A, bottom) and anti-correlation with adaptation time (Figure 2 – figure supplement 1A, top)[4]. Here, this variation arises solely from variability in protein levels, as these were the sole quantities that were varied between cells within a population. Although this variation exists within the population, the population average CW is constrained within a functional range even when the mean level of proteins is globally upregulated (Figure 2 – figure supplement 1B) – experimentally this was done by inducing the expression of a master transcriptional regulator[7], here we multiplied **P**$_0$ by the appropriate factor. This conservation shows that our model recapitulates and resolves a fundamental unexplained dichotomy in the chemotaxis pathway: population variability around the average is possible in addition to high robustness of the population average. Hence the pathway is sensitive to molecular noise at the single cell level, but robust at the population level. Finally, simulations of the population model to reproduce the experiment by Shimizu et al.[42] show close agreement (Figure 2 – figure supplement 1C), notably fitting the nonlinear behavior at low and high receptor activity levels without using a piecewise model or higher exponents in the methylation equation.

Phenotypic model of chemotaxis

The stochastic molecular model described above and its linearization specifies the stochastic behavior of the single cell in a given environment as a function of its biochemical parameters (e.g. reaction rates) and protein concentrations. In the following, we define the key phenotypic parameters of the system, adaptation time, clockwise bias, and the dynamic range of CheY-P levels. We derive these quantities directly from the molecular model as a function of protein levels.

Consider the small changes in ligand concentration experienced by a cell moving in a gradient. In this linear regime, small perturbations in receptor activity around the mean steady-state value $a_0$ (given by the steady state of Eq. (7)) will arise from either intrinsic molecular noise or from the external stimuli. The relaxation time of the system $\tau$ (Eq. (13)) is determined by the time scale of methylation and demethylation, which are slow relative to all other reactions in the system (SI Table 1) and may therefore be considered at steady state relative to methylation. Under these conditions, we may construct a simplified version of the above pathway model with only a single SDE to describe the methylation dynamics. The end result is a phenotypic model specified only by the mean activity at steady-state $a_0$, the relaxation time scale $\tau$, and the total CheY concentration $Y_{Tot}$ that controls the maximum level of CheY-P that the cells can reach and therefore the dynamic range of the response regulator.

In the phenotypic model, the dynamics of the mean receptor methylation level $m$ are described by[17]:

$$\frac{dm}{dt} = -\frac{1}{\tau}(m - m_0(L)) + \eta_m(t). \tag{19}$$

For a given ligand concentration $L$, $m_0(L)$ is the methylation level at which receptor activity is equal to its mean adapted level $a_0$. Therefore $m_0$ satisfies $a(m_0, L) = a_0$ with $a$ given by Eq. (4) above and $a_0$ given by the steady state of Eq. (7). The white noise source $\eta_m$ is identical to that in Eq. (6) with intensity derived from $\tau$ according to Eqs. (13-16).

The phosphorylation reactions described by Eqs. (8-10) are much faster than the methylation and demethylation reactions (Eq. (7)) that govern the slow adaptation of the cell and therefore are calculated using a steady-state approximation as in previous studies[17, 23, 75]. Since the concentration of total CheB is small relative to total CheY ($B_{Tot}/Y_{Tot} \ll 1$) and the rate of CheB phosphorylation is lower than the rate of CheY phosphorylation, the effect of CheB phosphorylation in Eq. (8) can be safely neglected. Solving Eqs. (8) and (10) then yields the following relationship between CheY-P concentration and the kinase activity $a$:

$$Y_P(a) = \frac{1}{2}\left( a\frac{a_p}{a_y} + Y_{tot} + a\alpha - \sqrt{-4aY_{tot}\alpha + \left(a\frac{a_p}{a_y} + Y_{tot} + a\alpha\right)^2} \right), \qquad (20)$$

where $\alpha = \frac{A_{Tot} a_p}{Z_{Tot} d_z}$. Phosphotransfer from CheA to CheY is rapid. Consequently, if $Y_{tot}$ is sufficiently large that $a_p = a_y Y_{tot}$, then equation (20) reduces to $Y_p(t) \cong \alpha Y_{tot} a(t)$. This linear relationship has been exploited by researchers using CheY–CheZ FRET as a read-out of kinase activity[76]. Thus, for large $Y_{tot}$, the relationship between kinase activity $a$ and CheY-P concentration is nearly linear with slope $\alpha$. Also see below under section "*Simulating Performance of Phenotypes.*"

In summary, we combine the phenotypic model Eqs. (19-20) with the MWC receptor model Eq. (4-5) and the flagellar motor switching model Eq. (1-3) to produce a simplified model of the bacterial chemotaxis system in the linear regime.

Using this model, an individual cell is fully specified by the three parameters: clockwise bias, adaptation time, and the dynamic range of the response regulator CheY-P:

1) The <u>clockwise bias</u> can be obtained from the molecular model (Eqs. (6-10)) at steady state using the protein levels ($A_{tot}$, $T_{tot}$, …) and biochemical parameters ($k_r$, $k_b$, …) to first obtain $a_0$ and $Y_{p,SS}$ and then by using Eq. (3) to solve for the steady-state clockwise bias as a function of $Y_{p,SS}$.

2) The <u>adaptation time</u> can be obtained from Eqs. (12-14), which depend on the molecular model (Eqs. (6-10)) that is parameterized by the protein levels ($A_{tot}$, $T_{tot}$, …) and biochemical parameters ($k_r$, $k_b$, …). That value of adaptation time also directly sets the adaptation time in the phenotypic model described in Eq. (19).

3) The <u>dynamic range</u> of the response regulator CheY-P is defined as $Y_p(a=1)$ in Eq. (20) and is determined by the total number of CheY molecules in the cell, $Y_{tot}$. For large values of $Y_{tot}$ the response regulator activity is linear with that of the kinase and therefore the maximum level of $Y_p$ the cell can adopt is $\alpha$. For lower values of $Y_{tot}$, the total amount of CheY proteins in the cells becomes limiting and the dynamic range of CheY-P diminishes proportionally to $Y_{tot}$.

The values of all parameters used in this study are given in Supplementary File 1 – Parameter Values.

Model parameter summary

Collectively our model therefore consists of the three classes of parameters:

- Biochemical parameters of the signaling network ($k_r$, $k_b$, $K_r$, $K_b$, $a_p$, $a_y$, $d_z$, $d_b$, $a_b$) represent the physical kinetics of the proteins' enzymatic actions. In this paper, these parameters are fixed for all populations in all cases because we assume neither the genes nor the pathway topology changes.
- Population parameters of the gene expression model ($\mathbf{P}_0$, $\eta$, $\omega$) represent the genetic architecture (i.e. operons, promoters, and RBSs) of the chemotaxis genes shared by all individuals in the clonal population. In this paper, these parameters can vary at the population level (such as in Figure 6 and the population optimization for Figure 7) but are assumed to be the same within populations. Their role here is to determine the distribution of protein levels among individuals within a given population.
- Phenotypic parameters of the cell (adaptation time, clockwise bias, dynamic range of the response regulator) control the dynamical behavior of the individual cell. These vary from cell to cell and are determined by the combination of the individual levels of protein generated by the populations' noisy gene expression parameters and the biochemical signaling network as described in Methods section "*Phenotypic Model of Chemotaxis*" above. These parameters were also varied manually to perform the parameter scans of cell dynamics (See below).

Comparison of Molecular and Phenotypic Models

We tested the agreement between the molecular model specified by Eqs. (1-10) and the phenotypic model specified by Eqs. (1-5,19-20) in two types of simulations. First we performed deterministic simulations of immobilized cells being exposed to 50 s square pulses of attractant and compared the time traces of CheY-P output of the system (Figure 2 – figure supplement 2) for a wide range of adaptation time, clockwise bias, and $Y_{Tot}$. The two traces lie on top of each other, demonstrating agreement.

*Stochastic simulations of the model in ecological challenges*

Environment definitions

We simulated cell trajectories using the phenotypic model[17] in 3-D environments in which methyl-aspartate was diffusing. The sources of methyl-aspartate were spherical and diffusion was modeled as a 1-D process with central symmetry extending from the center of the source, described by:

$$\frac{\partial L}{\partial t} = D \frac{1}{r^2} \frac{\partial}{\partial r}\left( r^2 \frac{\partial L}{\partial r} \right) \tag{21}$$

where $L$ is the ligand concentration at radius $r$ in µm from the center of the source and $D$ is the diffusion coefficient of methyl-aspartate.

In the foraging simulations, the only boundary condition was that $L$ goes to zero as $r$ goes to infinity. The source simply diffused from its spherical initial condition, which was given by:

$$L(r, t=0) = \begin{cases} L_1 & r \leq R \\ 0 & r > R \end{cases}, \tag{22}$$

where $L_1$ was the initial source concentration and $R$ was the radius of the source. The solution of the gradient in space and time becomes

$$L(r,t) = L_1 \frac{1}{r}\sqrt{\frac{Dt}{\pi}}\left(e^{-\alpha^2} - e^{-\beta^2}\right) + L_1 \frac{1}{2}\left(\text{erf}(\alpha) - \text{erf}(\beta)\right), \tag{24}$$

where $\alpha = \frac{r+R}{\sqrt{4Dt}}$ and $\beta = \frac{r-R}{\sqrt{4Dt}}$.

In the colonization simulations, there was an additional boundary condition to describe the persistence of the source as a permanent non-depleting zone:

$$L(r \leq R, t) = L_1, \tag{24}$$

resulting in the solution

$$L(r,t) = \begin{cases} L_1 & r \leq R \\ L_1 \frac{R}{r}\text{erfc}\left(\frac{r-R}{\sqrt{2Dt}}\right) & r > R \end{cases}. \tag{25}$$

Simulating performance of phenotypes

To construct heatmaps, adaptation time was varied directly, and the internal parameters used to vary the clockwise bias and dynamic range were $Y_{p,SS}$ and $Y_{tot}$, respectively. Adaptation time was scanned in 25 log-spaced steps over the interval [1,300] s. $Y_{p,SS}$ was scanned in 25 log-spaced steps over the interval [1.2,4] μM with an additional point at 0 μM. $Y_{tot}$ was scanned in 10 log-spaced steps over the interval [820,82000] molecules/cell. For each combination of parameters, 6000–30000 replicates were simulated; more replicates were used for farther away sources.

To calculate foraging performance, individual cell replicates accumulated ligand along their trajectories during the simulation. At each time point, $kL(t)dt$ was added to their total, where $k$ is the uptake rate in μmol/μM/s, $L(t)$ is the concentration of ligand at the current cell position and time, and $dt$ is the simulation timestep in seconds. Nutrient accumulation was cut-off after a certain time; for the cases in the main figures, this was 13 minutes – characterization of the effect other times are described below. For each phenotype, the performance was the average ligand accumulation of all replicates (Figure 2 – figure supplement 3A).

To calculate colonization performance, each cell replicate was first awarded an arrival time in minutes upon its first crossing of $r = R$. If it failed to cross in 15 min, its arrival time was infinity. Since we define performance as a quantity to be maximized rather than minimized, the performance of each replicate was defined as the reciprocal of the arrival time in min$^{-1}$; infinite arrival time resulted in zero performance. The replicates of each phenotype were then averaged together to get the performance as a function of phenotype (Figure 2 – figure supplement 3B).

Replicate cells began at a single point on the spherical shell $r = r_0$, where $r_0$ is the distance away that the source appears to the cells. Simulations in both tasks were performed for 15 minutes for several values of $r_0$ (Figure 2 – figure supplement 3AB). For all subsequent analysis, the heatmaps were smoothed and resampled at higher resolution in phenotype space (Figure 2 – figure supplement 3CD). Performance was first smoothed over the original cell parameter mesh using a Gaussian kernel. A resampled parameter mesh was created on a 200 x 200 grid within the existing range of cell parameters. The performance values at the new grid points were interpolated from the smoothed performance heatmaps using spline interpolation. The same procedure was used for the fitness landscapes described later, but importantly all fitness calculations were performed on original unsmoothed raw data, then smoothing was applied, so we did not smooth twice.

To find the specialist in each environments, the top performing phenotype on the high-resolution parameter grid was screened out (Figure 2 – figure supplement 3CD, diamonds). To plot the Pareto front for two environments, we scattered the performance of all values of adaptation time and $Y_{p,SS}$ (for a given $Y_{tot}$) on the high-resolution grid to obtain a cloud in chemotactic performance space. We then screened out phenotypes which were *not outperformed* in environment 1 and *not outperformed* in environment 2 to form the Pareto front.

Changing time limits was explicitly considered in the Results for the colonization case. For foraging, reducing the time limit has little effect on sources that are nearby, but for far sources, it limits the success of high clockwise bias dramatically, exaggerating the trends observed in the far cases (Figure 2 – figure supplement 5). In this regime, time is a key parameter since the cells must explore to reach the front of the source before foraging it appreciably. As such, limiting the time has a disproportionate effect on these phenotypes.

As a result of these disproportionate effects in far environments, trade-offs become stronger when time is restricted (Figure 2 – figure supplement 6). For the same four examples of two-environment near–far trade-offs, reducing the time consistently increases the curvature of the Pareto front. When 3 minutes are given, the transition of trade-offs from weak to strong occurs at a much more similar pair of environments. Therefore, if the environment has frequent turnover events, for example caused by turbulence that equilibrates the gradient, it strengthens trade-offs substantially.

In the cases shown in the Results section, $L_1$ was 100 mM for foraging and 10 mM for colonization. If $L_1$ is reduced (Figure 2 – figure supplement 7), the colonization case is relatively unaffected, except that the trend in farther sources become exaggerated because cells have to explore longer before finding a detectable signal. There are more substantial changes to the foraging case. The for farther sources especially, the preferred phenotypes switch to having high clockwise bias. In these cases, exploration reduces the chances of the cells to see ligand because they become too spread; rather, staying in one place and waiting for the diffusing nutrient front to arrive becomes the preferred strategy.

As we derived in equation (20), the dynamic range of CheY-P depends on $Y_{tot}$, which sets the asymptotic value of CheY-P. In cells with low $Y_{tot}$, phosphotransfer is hindered, reducing information transfer from the kinase to the motor and thus deteriorating performance. Cell performance is limited by low $Y_{tot}$, but once it is high enough to reach the linear regime between kinase activity and CheY-P concentration, additional CheY does not add much benefit since the dynamic range of CheY-P activity will then become limited by the number of kinases.

We see in our simulations (Figure 2 – figure supplement 8) that, above about $Y_{tot}$ = 10591 molecules/cell, the performance does not appreciably change because this condition of linearity is met. From this, we conclude that there is no trade-off on $Y_{tot}$ apart from the cost of protein synthesis, and that cells should express enough CheY to reach the Pareto front. Beyond that, there is minimal increase in performance. Since the Pareto front represents the outer bound of performance, in Figures 4 and 5 we used $Y_{tot}$ = 29469 mol./cell for all cells; the results do not change significantly if the next higher or lower levels of $Y_{tot}$ are used instead.

Calculating fitness from performance

Fitness was assigned based on performance via a selection function. The fitness of each individual simulation trajectory was calculated, then all trajectories of a given phenotype were averaged together to produce the fitness of a given phenotype. This is clearly distinct from calculating the fitness of each phenotype's average performance. We used this procedure to create fitness landscapes which were then smoothed and resampled exactly as we did with the performance heatmaps.

Fitness was calculated on a single-cell (i.e. single-replicate) basis. In the foraging case, our metabolic formula was $f = \left[1 + (K/N_{col.})^n\right]^{-1}$, where $K$ is the amount of nutrition required for survival and $n$ is the dependency; for colonization, our time-limit model was $f = H(T_L - T_{arr.})$, where $T_L$ is the time limit, and $H$ is the Heaviside step function.

In addition to the fitness functions described in the Results section, we also tested two additional cases for increased generality (Figure 5 – figure supplement 1). For the foraging case, different levels of nutrition may be associated with discrete transitions to different physiological states. If the nutrition is below a survival threshold $T_{survive}$, the individual dies, resulting in an outcome of 0 to signify no progeny. If the nutrition is above a higher division threshold $T_{divide}$, the individual gives rise to 2 progeny. Nutrition in between the two thresholds results in survival of the individual, or an outcome of 1 progeny. This model can be written as: $f = H(N_{col} - T_{survive}) + H(N_{col} - T_{divide})$ (Figure 5 – figure supplement 1A).

Similar to the case of the continuous, probabilistic model of survival (Figure 4A-C), lower thresholds (Figure 5 – figure supplement 1A, blue line) result in a neutral performance trade-off (Figure 4B) giving rise to a weak fitness tradeoff (Figure 5 – figure supplement 1B), whereas higher thresholds (Figure 5 – figure supplement 1A, red line) transform the same performance trade-off into a strong fitness trade-off (Figure 5 – figure supplement 1C)

We can consider the effect of a continuous, probabilistic model of selection applied to colonization. In our example in the Results section (Figure 5D-F), colonization was an all-or-nothing deterministic outcome depending on whether an individual arrived within the time constraint. If colonization depends on arrival time within a certain time limit $K$, but that dependency $n$ is not absolute (i.e. infinite), this could be described by a sigmoidal function like the Hill equation: $f(V) = 1 - \left(1 + \dfrac{K^n}{T_{arr.}^n}\right)^{-1}$ (Figure 5 – figure supplement 1D).

When $n$ is very high (Figure 5 – figure supplement 1D, red line), the result is very similar to that in the discrete transition model: a neutral trade-off can be converted into strong trade-off when the threshold is low (Figure 5 – figure supplement 1F). On the other hand, if the probability of colonization depends less strongly on the arrival time (Figure 5 – figure supplement 1D, blue line), a weak trade-off may result from the same underlying performance trade-off (Figure 5 – figure supplement 1E).

For each selection function and for each environment, the approach described above creates a lookup table for fitness as a function of phenotype. In order to calculate population fitness below, we must calculate the fitness of individual cells that are initially defined only by their levels of chemotaxis proteins. In order to find $Y_{p,SS}$, we solve the system of Eqs. (5,7-10). To find adaptation time we use the definitions in Eqs. (11-13). $Y_{Tot}$ is given as one of the protein levels.

Given these three phenotypic parameters, we interpolate on the lookup table for any combination of task, environment, and selection function to give the corresponding fitness of that phenotype.

*Optimization of gene expression parameters under trade-offs*

To optimize population fitness, we first defined a general expression for population fitness beginning with the fitness of a single phenotype. Chemotaxis is non-deterministic, hence, in each environment $g$, an individual phenotype $\vec{x}$ had a distribution of performance $V$, or $p(V|\vec{x},g)$, where $\vec{x}$ is a vector of adaptation time, clockwise bias, and CheY-P dynamic range. Fitness was a function of single-cell performance $f(V)$. To calculate the fitness of a phenotype in a given environment, we took the expected value of its fitness over its distribution of performance $\langle f \rangle_{\vec{x},g} = \int f(V) p(V|\vec{x},g) dV$. This should not be confused with the fitness of the average performance.

We assume for simplicity that populations encounter challenges sequentially, all cells in the population experience each challenge simultaneously and in the same way, and populations must survive through all environments. Hence, within a given environment, a population consisting of many cells with different phenotypes has fitness equal to the average of its constituent cells $\langle f \rangle_{P,g} = \int P(\vec{x}) \langle f \rangle_{\vec{x},g} dx$, where $P(\vec{x})$ is the population distribution of phenotypes. Following this, population fitness from one environment to the next is multiplicative. In the long term this results in a geometric mean across environments, weighted by the probability of encountering each environment:

$$F = \exp\left(\int \log\left(\int \langle f \rangle_{P,g}\right) h(g) dg\right), \quad (26)$$

where $h(g)$ is the distribution of environments. This formula is consistent with previous derivations[22] but has been extended to include stochastic performance of individual cells and a distinction between fitness and performance.

While equation (26) provides a general solution, in the specific cases analyzed in this study, the populations consist of a finite number of different phenotypes, there are a finite number of discrete simulation replicates, and for simplicity we show cases that compare two discrete environments $g_1$ and $g_2$ with occurrence probability $h$ and $(1 - h)$. As such, the discrete calculation of population fitness becomes:

$$F = \left(\sum_{\xi=1}^{N_{pop}} \frac{f_{\xi,g_1}}{N_{pop}}\right)^h \left(\sum_{\xi=1}^{N_{pop}} \frac{f_{\xi,g_2}}{N_{pop}}\right)^{(1-h)}, \quad (27)$$

where $\xi$ indexes the cells in the population, $N_{pop}$ is the number of cells in the population, and $f_{\xi,g}$ is the fitness of the phenotype of cell $\xi$ in environment $g$ determined using a look-up table constructed from simulation data as described above.

The trade-off problem itself is thus parameterized by: $h$, $g_1$, $g_2$, and the form and parameters of $f(V)$ that gave rise to the look-up table. In the case of foraging these are the nutritional requirement *K* and the dependency *n*; for colonization there is only the time limit $T_L$. These we collectively call the trade-off parameters.

The population gene expression parameters generate a list of individuals with different phenotypes as described above. We can optimize the fitness of the population as a whole (Figure 7) by first calculating population fitness *F* (Eq. (27)) for a set of trade-off parameters. We then used MATLAB's pattern search optimization function on the population fitness formula, allowing only the gene expression parameters $\mathbf{P_0}$, $\eta$, and $\omega$ to vary, but not the trade-off parameters, the biochemical parameters, or any other parameters. The constraints on these parameters are described below, and *h* was 0.8. From this we obtained the optimized population parameters for strong and weak trade-offs (performed separately). For each type of ecological task, the strong and weak trade-offs are between the same pair of near and far environments, with the same form of selection function, but each has a different set of selection function parameters.

Since there is always some irreducible noise in biology, we used experimental observations to provide lower bounds for the noise parameters in our model. For a limit on the intrinsic component, we took the wildtype level of intrinsic noise, which we obtained by fitting the model to wildtype data (described above). Multiple studies have described the advantage of reduced intrinsic noise in chemotaxis, so we assume wildtype cells are likely to be functioning at or near the minimum intrinsic noise. In order to apply this constraint, we ensure that the intrinsic noise scaling parameter and mean protein levels are constrained within the optimization algorithm such that the condition $\eta \mathbf{P_0} \geq \eta^{wt} \mathbf{P_0}^{wt}$ is maintained.

There is also a lower bound on the minimum total protein noise, defined as the coefficient of variation squared, measured in single *E. coli* cells to be about 0.09 for proteins with a mean expression level of above 100 copies per cell[54]. This constraint in practice acts more on extrinsic noise than on intrinsic noise since in our case the latter is typically fairly low. To enforce this constraint computationally, we ensure that $\mathbf{P_0}$, $\eta$, and $\omega$ of Eq. (18) are chosen by the optimization such that the squared coefficient of variation of every protein is above 0.09. This typically has the effect of keeping $\omega$ above about 0.09, depending on $\mathbf{P_0}$ and $\eta$. Increases in global expression levels of up to approximately 3 fold are observed for different strains and growth media[50], and using mutations in *flgM*, increases up to 7 fold are possible[7]. Hence, we set our upper limit of mean expression levels at 5 fold to work within that range.

**Acknowledgements**

We thank P. Cluzel, T. S. Shimizu, P. Turner, G. Wagner, and S. Zucker for valuable discussion on the manuscript. This research was supported by the James S. McDonnell Foundation (award no. 220020224), the Paul Allen foundation (award no. 11562), and the National Institute of Health (grant no. 1R01GM106189). Simulations were performed at the Yale University Faculty of Arts and Sciences High Performance Computing Center, which is partially funded by the National Science Foundation under grant #CNS 08-21132

**Figure legends**

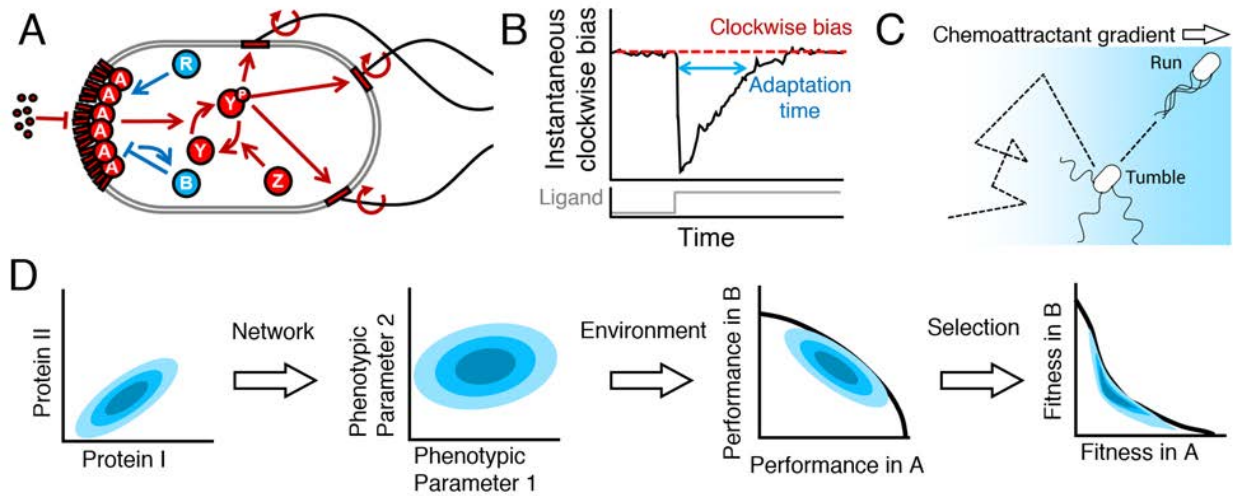

**Figure 1.**
**From proteins to fitness.**
**A**. The cell receives extracellular ligand signals through transmembrane receptors. Changes in signal are rapidly communicated to the flagellar motors through the kinase CheA and response regulator CheY. CheZ opposes the kinase activity of CheA. At a slower timescale, the activity of the receptor complex physiologically adapts to its steady-state activity through the antagonistic actions of CheR and CheB. **B**. Cartoon diagram of the response of the system to transient step-stimulus and definition of the key phenotypic parameters of the system. Without stimulation, the system has a steady-state clockwise bias, or fraction of time spent with motors in the clockwise state that results in tumbling. Upon stimulus with a step, CheY activity and therefore clockwise bias drops and the cell starts running more, then slowly adapts back to the steady-state with a characteristic timescale (adaptation time). The steady-state clockwise bias and adaptation time are tuned by the concentrations of proteins in A. **C**. Cells explore their environment by alternating between straight runs and direction-changing tumbles. When cells sense that they are traveling up a concentration gradient, they suppress tumbles to increase run length. Precisely how a cell navigates a gradient depends on its phenotypic parameters in B. **D**. From a single genotype, noise in gene expression leads to a distribution of proteins expression levels (blue shaded contours in protein space; left); network design determines how proteins quantities map onto phenotypic parameters (middle left); the performance of all possible phenotypic parameter values across environments will determine the outer boundary of performance space (middle right); selection bestows a fitness reward based on performance and will reshape the performance front into the Pareto front, which, for optimal fitness, the population distribution should be constrained to (right).

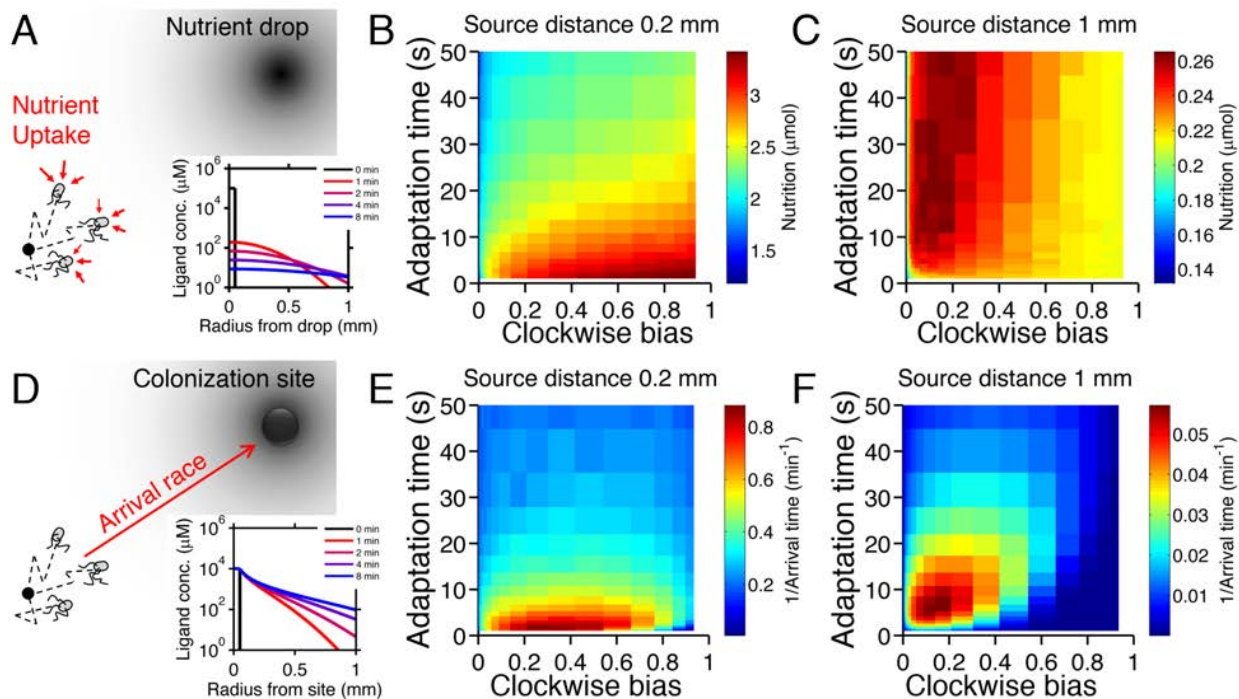

**Figure 2.**
**Performance of chemotactic phenotypes depends on environmental conditions.**
**A**. Cartoon diagram (not to scale) of the foraging challenge: cells navigating a 3-D time-varying gradient created by diffusion of a small spherical drop of nutrient 100 μm in diameter with diffusion coefficient of 550 μm$^2$/s and methyl-aspartate concentration of 100 mM. Inset: radial profile of the attractant concentration over time. **B**. Average nutrient collected by each phenotype (combination of clockwise bias and adaptation time) in environment in A over 8000 replicates per phenotype. Because we are investigating optimal phenotypes and CheY-P dynamic range does alter the results as long as it is sufficiently high (Figure 2 – figure supplement 7), results shown here use $Y_{Tot}$ = 13149 mol./cell. Clockwise bias and adaptation time were sample in log-spaced bins. Cells start near to the source (0.2 mm from its center), and are allowed to swim for 13 minutes while accumulating a small fraction of the nutrient they sense. **C**. Same as B except that cells start farther away from the source (1 mm from its center) and 14000 replicates per phenotype were used. **D–F**. Similar to A–C but the environment consists of a colonization challenge: diffusion of ligand out of a spherical non-depleting source representing a colonization site; source methyl-aspartate concentration was 10 mM. Rather than nutrient collection, performance (E, F) was quantified as the reciprocal of the arrival time at the source averaged over all replicates (9000 and 36000 for E and F respectively) with a maximum time allotted of 15 minutes.

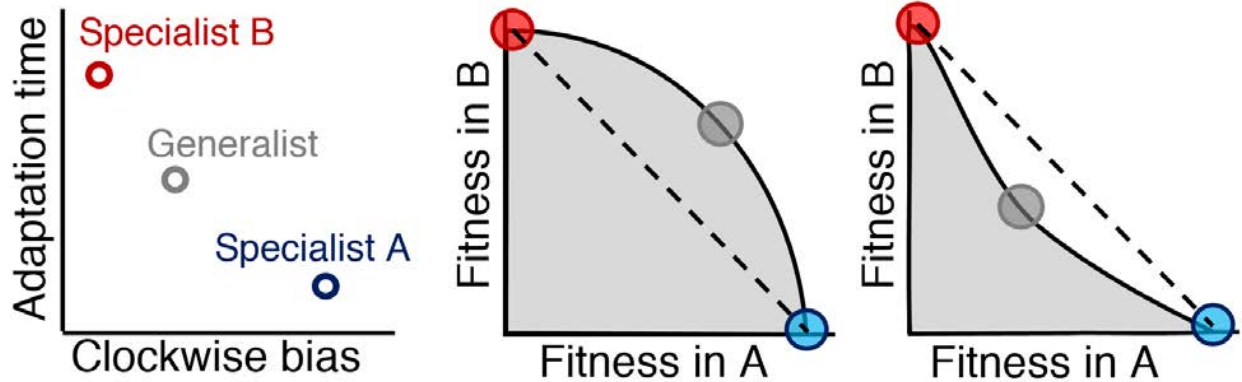

**Figure 3.**
**Relationship between Pareto front shape and population strategy.**
Left: Two environments, A and B, select for different optimal phenotypes, specialist A and specialist B (blue and red circles). The generalist phenotype (gray circle) performs well, but not optimally, in both environments. Middle and right: Trade-off plots. Gray region: fitness set composed of the fitness of all possible phenotypes in each environment; Black line: Pareto front of most competitive phenotypes; Dashed line: fitness of mixed populations of specialists; Circles: fitness of phenotypes corresponding the circles in the left plot. Middle: In a weak trade-off (convex front), the optimal population distribution will consist purely of a generalist phenotype that lies on the Pareto front. Right: In a strong trade-off (concave front), the optimal population will be distributed between the specialists for the different environments. Here, the fitness of a mixed population of specialists (dashed line), exceeds that of the generalist in both environments.

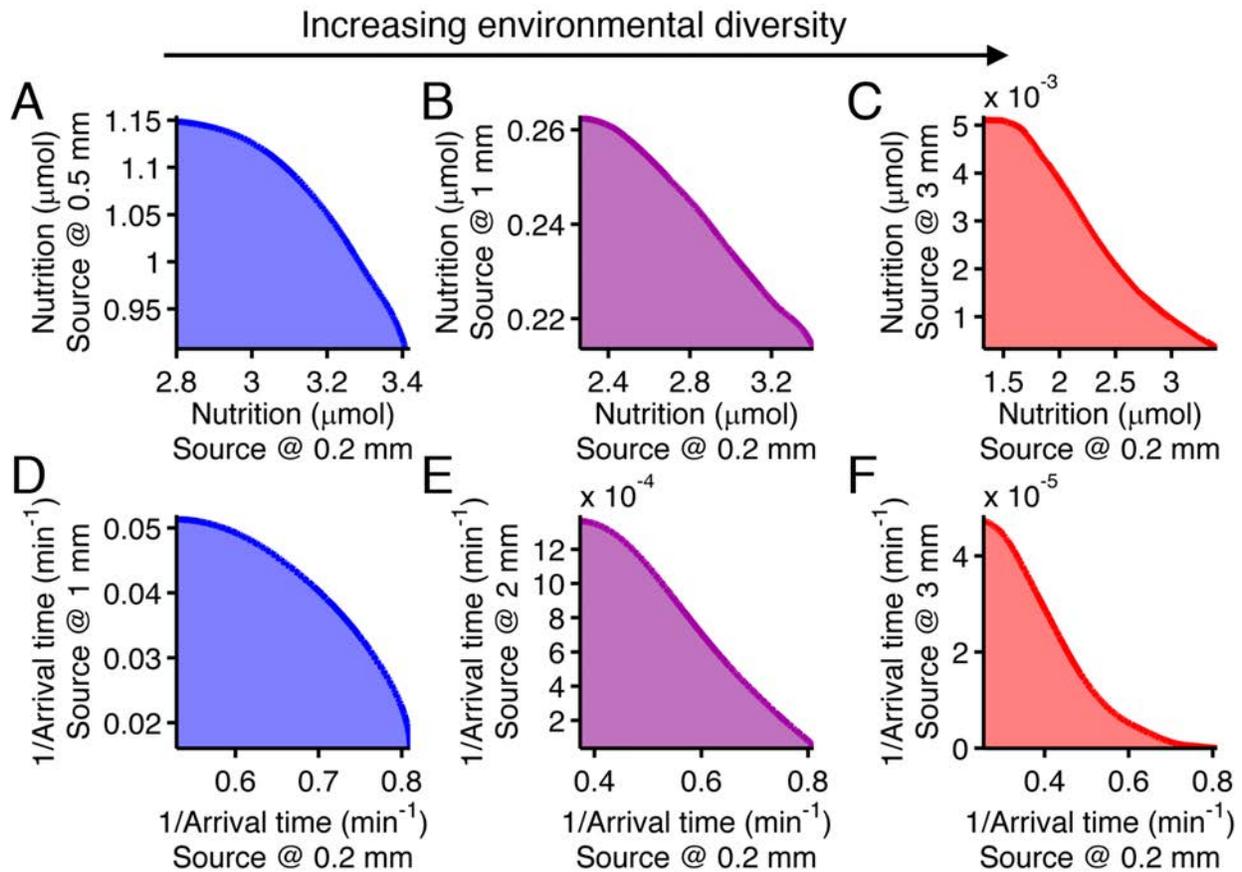

**Figure 4.**
**Performance trade-offs in *E. coli* chemotaxis.**
Ecological chemotaxis tasks pose trade-off problems for *E. coli* that become strong when environmental variation is high. (**A–C**). Trade-off plot between nutrient accumulation when starting near and when starting far from a source. Plotting the performance of all possible clockwise bias and adaptation time combinations in both near and far cases (colored region) reveals the strength of the trade-off in the curvature of the front. As the disparity between starting distance becomes greater (left to right plots), the trade-off front goes from convex to concave, signifying a transition from weak to strong performance trade-offs. Source distances are indicated on axis labels. (**D–F**). Same as A–C but for the colonization challenge.

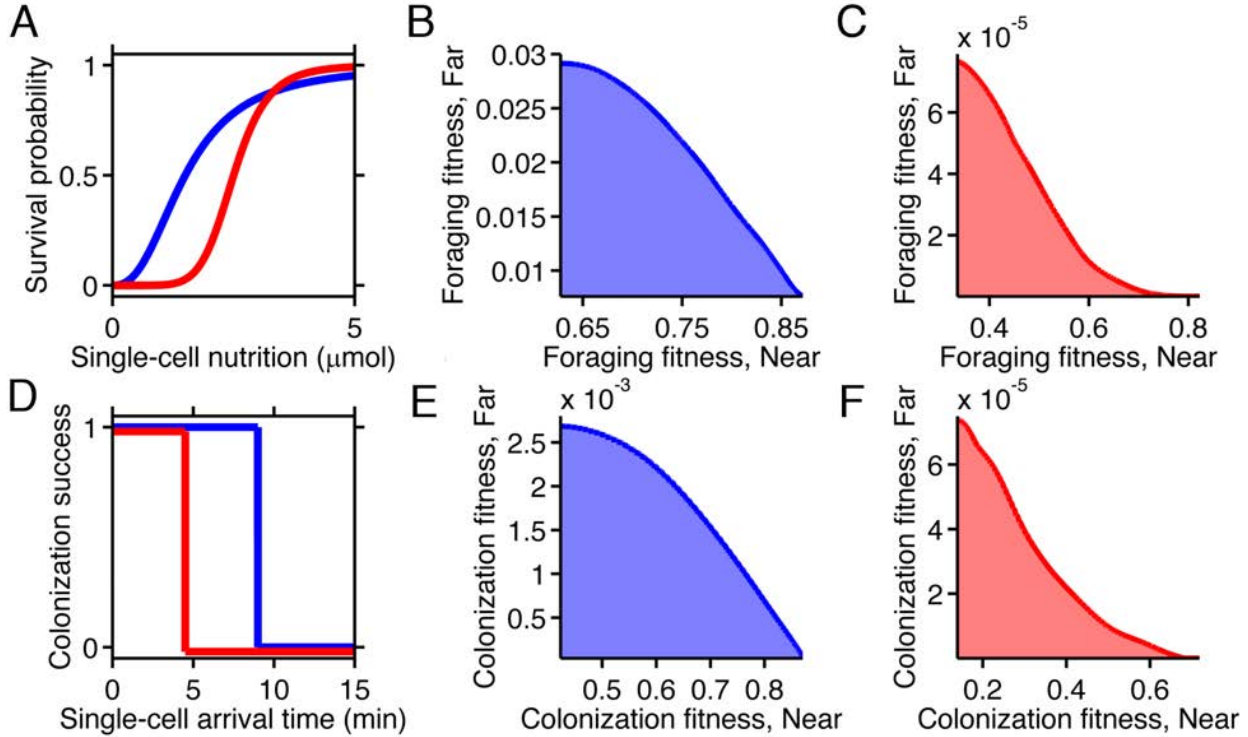

**Figure 5.**
**Selection can reshape trade-offs.**
**A**. Simple metabolic model of survival applied to the chemotactic foraging challenge. Each individual replicate is given a survival probability based on a Hill function of the nutrition they achieve from chemotaxis. For each phenotype, the foraging fitness is the average survival probability across replicates. The effect of more (red) and less (blue) stringent survival functions are compared. Transitional nutrition value: 1.5 μmol (blue), 2.5 μmol (red), Hill coefficient: 2.5 (blue), 7 (red). **B–C**. Beginning with the neutral foraging performance trade-off in Figure 4B, application of the survival model in A gives rise to either a weak (B) or strong (C) fitness trade-off, depending on whether the thresholds and steepness are low (blue curve in A) or high (red curve in A). **D**. Simple threshold model of survival applied to the chemotactic colonization challenge. Each individual replicate survives only if it arrives at the goal within the cut-off time. For each phenotype, the colonization fitness is the probability to colonize measured over all replicates. The effect of more (red) and less (blue) stringent survival functions are compared. Time threshold value: 5 min (blue), 1.5 min (red). **E–F**. Beginning with the neutral colonization trade-off in Figure 4E, application of the selection model in C gives rise to either a weak (E) or strong (F) fitness trade-off.

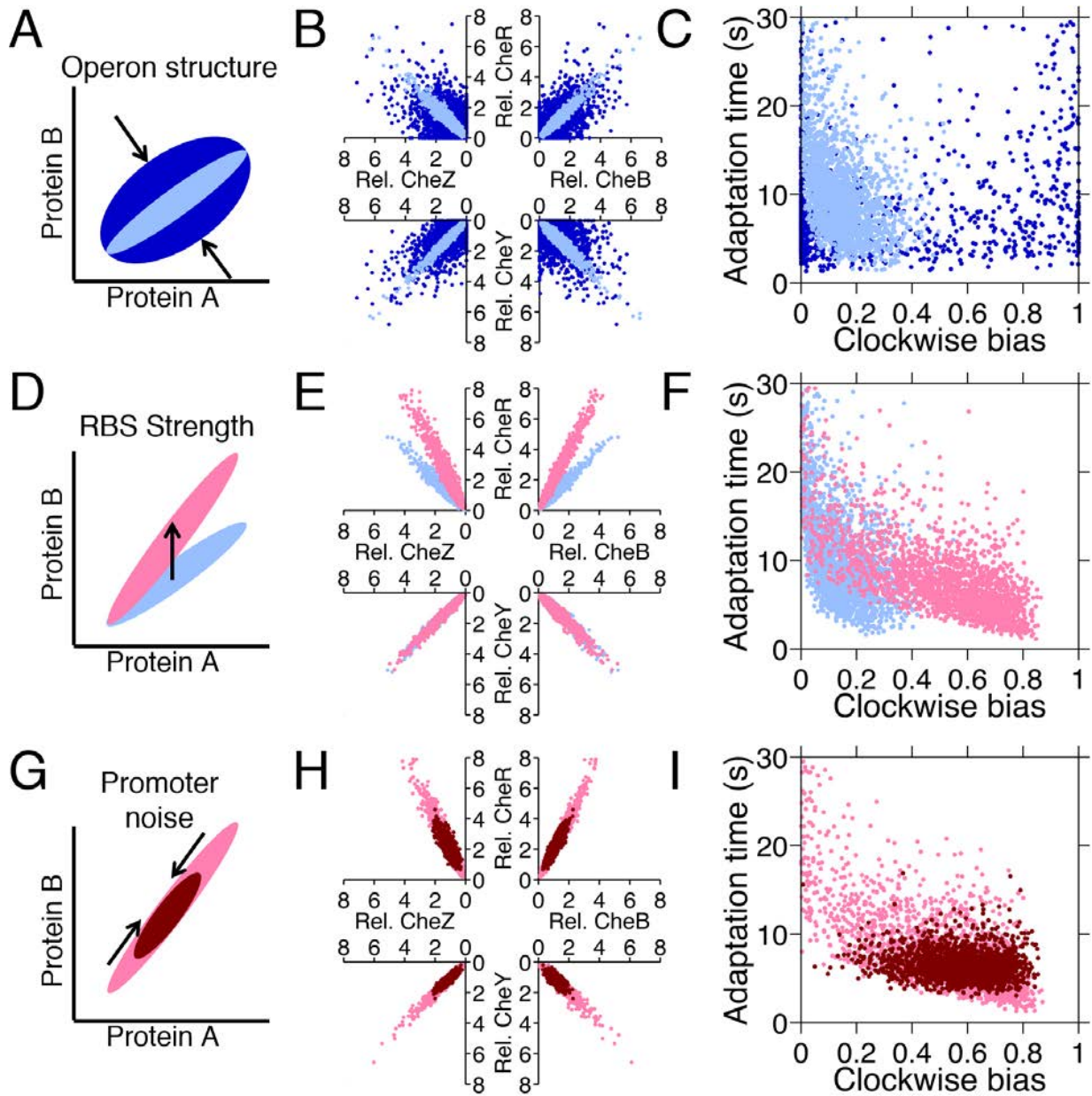

**Figure 6.**
**Genetic control of phenotypic diversity.**
**A**. Clustering genes on multicistronic operons constrains the ratios in protein abundance. **B**. Protein expression of core chemotaxis proteins CheRBYZ are shown relative to the mean level in wildtype cells. Two thousand cells are plotted. Light blue: mean levels of the proteins CheRBYZAW and receptors are equal to the mean levels in wildtype cells, which we take to be 140, 240, 8200, 3200, 6700, 6700, 15000 mol./cell, respectively[50]; the extrinsic noise scaling parameter, $\omega$, is 0.26 and the intrinsic noise scaling parameter, $\eta$, is 0.125, which are both equal to wildtype levels (Figure 2 —figure supplement 1). Dark blue: same but with $\omega = 0.8$, which is greater than wildtype level. Note the substantial variability around the mean even in the case of wildtype noise levels (light blue). **C**. Clockwise bias and adaptation time of individuals in A. **D**. Changes in the strength of individual RBSs will independently change the mean levels of individual proteins. **E–F**. Light blue: gene expression of cells with same population parameters as in A, light blue. Pink: mean levels of CheR changed to twice wildtype mean. **G**. Promoter sequences can be inherently more or less noisy, resulting in amplification or attenuation of the variability of total protein amounts without affecting protein ratios. **H–I**. Pink: gene expression of cells with same population parameters as in E, pink. Red: $\omega$ reduced from 0.26 to 0.1.

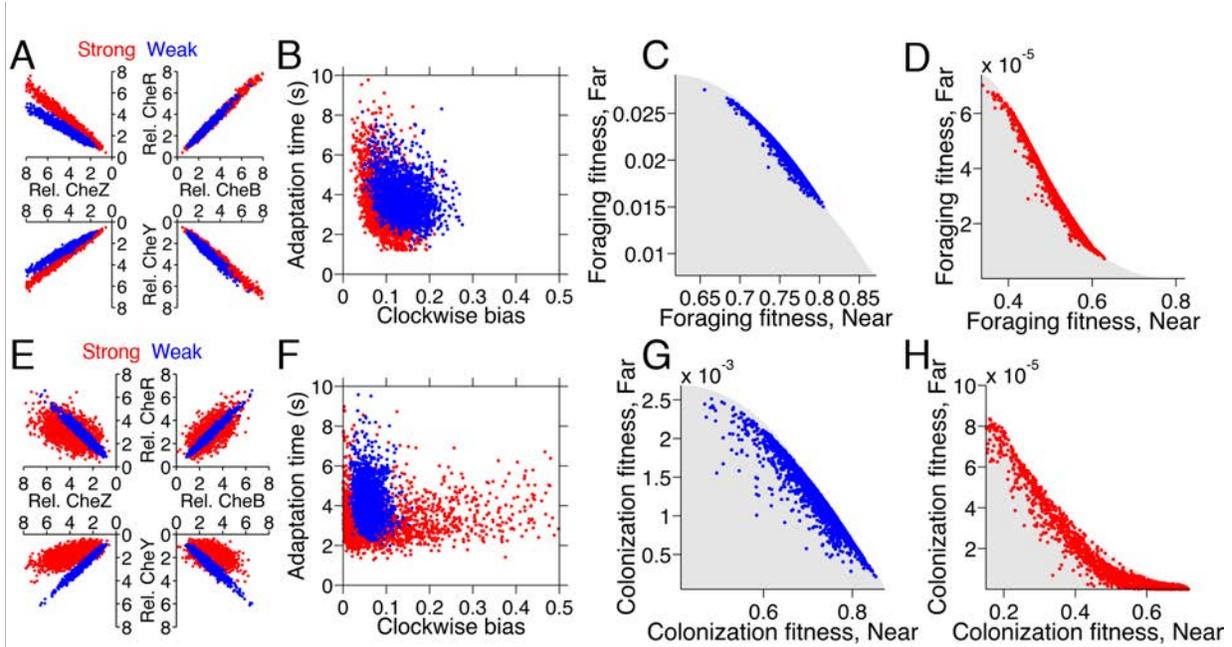

**Figure 7.**
**Optimization of gene expression noise reshapes population distributions to the Pareto front.**
Protein expression of populations were optimized for either weak or strong foraging fitness trade-offs (same trade-offs as in Figure 5B and C). For each population, 2000 individuals are plotted, protein expression shown relative to the mean level in wildtype cells. **A**. Gene expression parameters of the population optimized for the weak foraging trade-off: mean expression levels of CheRBYZAW and receptors relative to mean wildtype expression level were 2.54, 2.53, 2.50, 4.20, 4.50, 3.49, and 3.49 fold, respectively, with an *intrinsic* noise scaling parameter, $\eta$, of 0.051 and an *extrinsic* noise scaling parameter, $\omega$, of noise 0.128. For the strong foraging trade-off: mean CheRBYZAW and receptors relative to wildtype were 3.27, 3.27, 2.86, 3.83 4.17, 2.82, and 5.00 fold, respectively, with $\eta = 0.051$ and $\omega = 0.200$. **B**. Clockwise bias and adaptation time of individuals in A with the corresponding dot color. **C**. Fitness of the population that was optimized for the weak foraging trade-off (corresponding to blue dots in A and B). **D.** Same as C but for the population optimized for the strong foraging trade-off. **E–H**. Same as A–D for the colonization fitness trade-offs shown in Figure 5E and F. Population parameters optimized for weak colonization trade-off: mean CheRBYZAW and receptors levels relative to wildtype were 2.42, 2.52, 2.32, 2.50, 4.16, 3.25, and 5.00 fold, respectively, with $\eta = 0.055$ and $\omega = 0.126$. Population parameters optimized for strong colonization trade-off: mean CheRBYZAW and receptors levels were 2.90, 2.92, 1.71, 3.845, 2.25, 3.80, and 3.76 fold, respectively, with $\eta = 0.221$ and $\omega = 0.090$.

**Figure Supplement Legends**

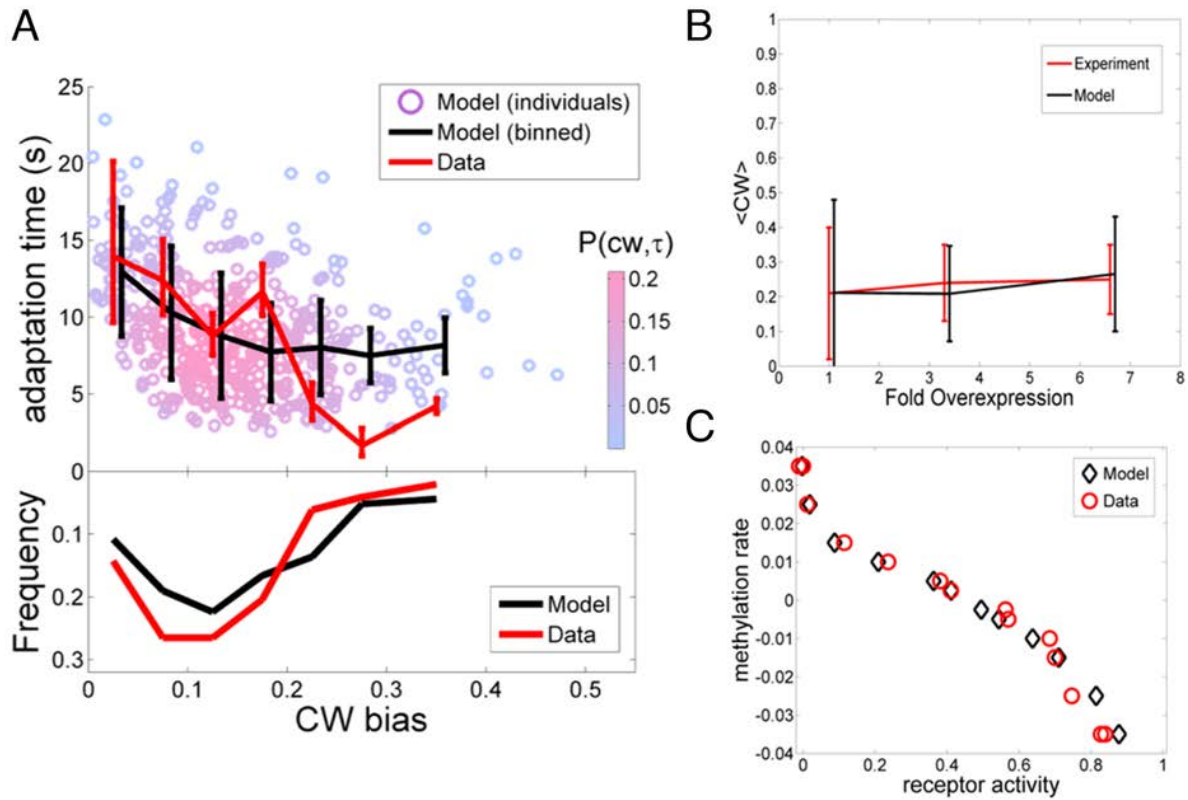

**Figure 2 – figure supplement 1.**
**Comparing the model to single cell and population averaged measurements.**
The same set of model parameter values is used for all the plots. **A**. Adaptation time and motor clockwise (CW) bias. Bottom: normalized histogram of motor clockwise bias in the population. Top: The mean and standard deviation of adaptation time in each bin of CW bias. Red lines: experimental data from ref. 4. Black lines: model. Circles: Individual cells from the model. Color: probability density. **B**. Population-averaged CW bias as a function of fold changes in mean expression level of all pathway proteins following ref. 7. Red: data from ref. 7. Black: model. **C**. Population-averaged methylation rate as a function of population-averaged receptor activity obtained by exposing cells to exponential ramps of methyl-aspartate as described in ref. 42. Red circles: data from ref. 42. Black: simulation of model.

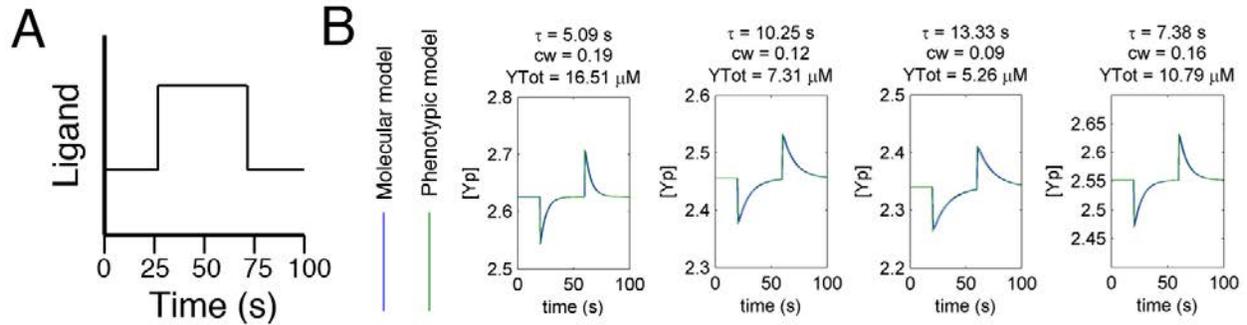

**Figure 2 – figure supplement 2.**
**Agreement between protein model and parametric dynamics model.**
**A**. Cartoon of step function of ligand delivered to immobilized cells in simulation to test response dynamics. **B**. Direct comparison of response of molecular model (blue) and phenotypic model (green) with the same parameters to stimulus of the form in A illustrating close agreement.

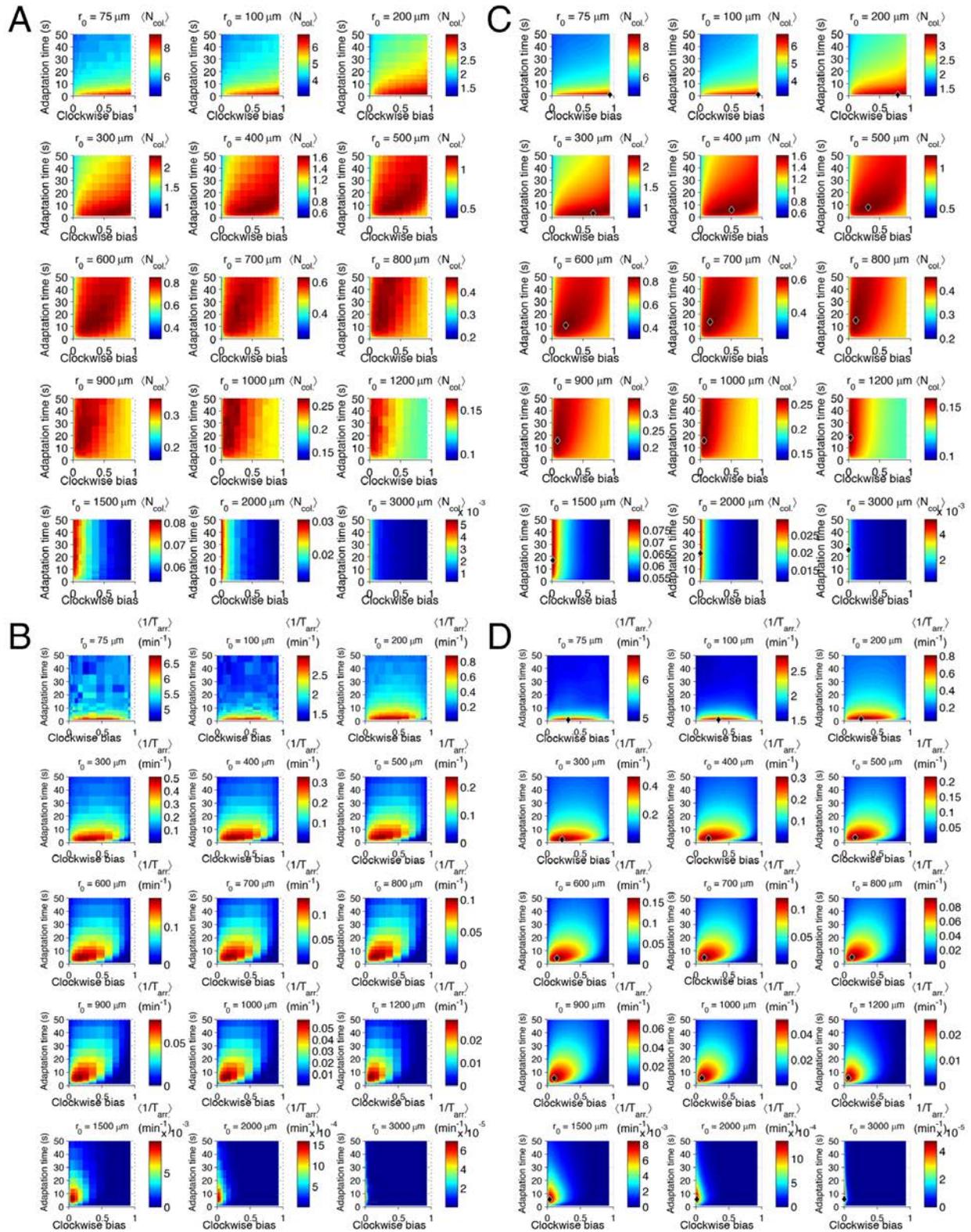

**Figure 2 – figure supplement 3.**
**Performance as a function of distance to source.**

**A**. Cells with various phenotypes were challenged to forage a source presented at varying distances, $r_0$ from 75μm to 3mm. Between 6000 and 30000 replicates were simulated for each phenotype. $\langle N_{col.} \rangle$: the average nutrient collected by all replicates of a given phenotype in μmol.
**B**. Same as A but for a colonization challenge; $\langle 1/T_{arr.} \rangle$: the average reciprocal-of-arrival-time of all of the replicates of a given phenotype in min$^{-1}$. **C**. Data in A smoothed with a Gaussian filter and resampled on a higher resolution grid of phenotypic parameters. Diamond: phenotype with highest performance. **D**. Same as C but with the data in B.

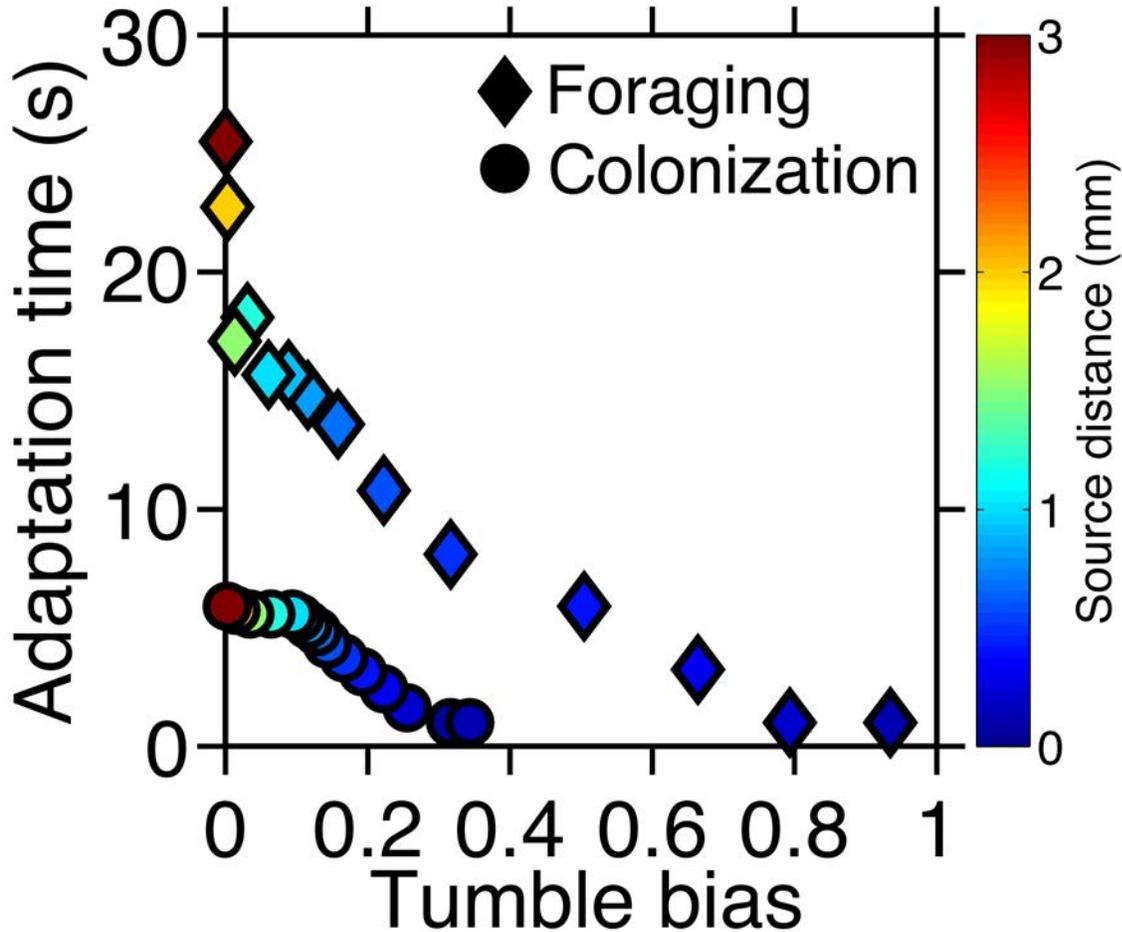

**Figure 2 – figure supplement 4.**
**Optimal phenotypes as a function of source distance.**
For each source distance and each task, the phenotype with highest performance was identified as shown in Figure 2 – figure supplement 3. The clockwise bias and adaptation time of these phenotypes are shown with the marker color corresponding to the distance to the source. Diamonds: foraging case. Circles: colonization case.

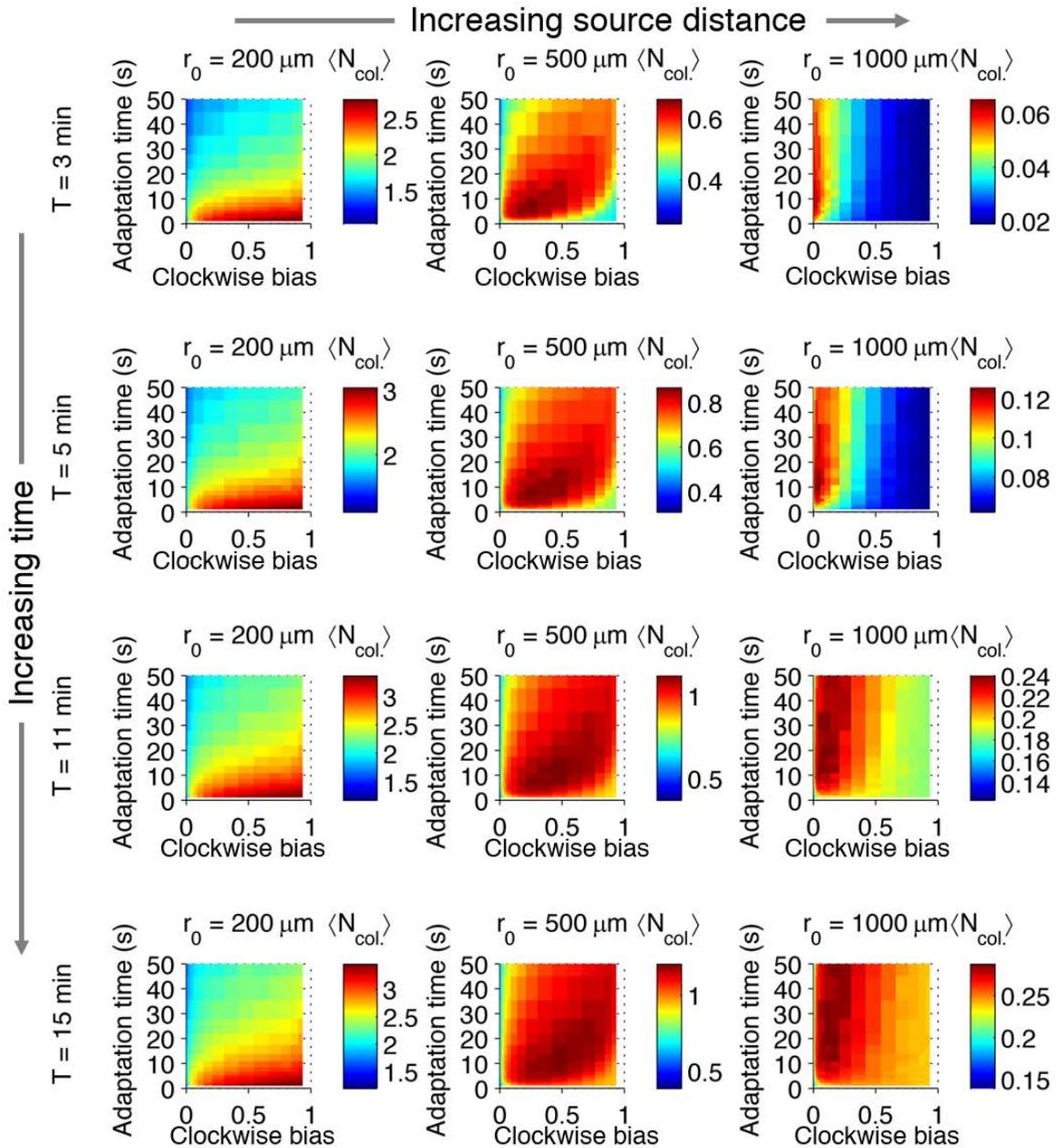

**Figure 2 – figure supplement 5.**
**Effect of time restrictions on foraging performance.**
Cells were challenged to forage sources that appeared at distances of 200, 5000, or 1000 μm away (columns from left to right). Different amounts of time were allotted to cells to accumulate ligand: 3 min, 5 min, 11 min, 15 min (rows from top to bottom). $\langle N_{col.} \rangle$ : the average nutrient collected by replicates of a given phenotype in μmol.

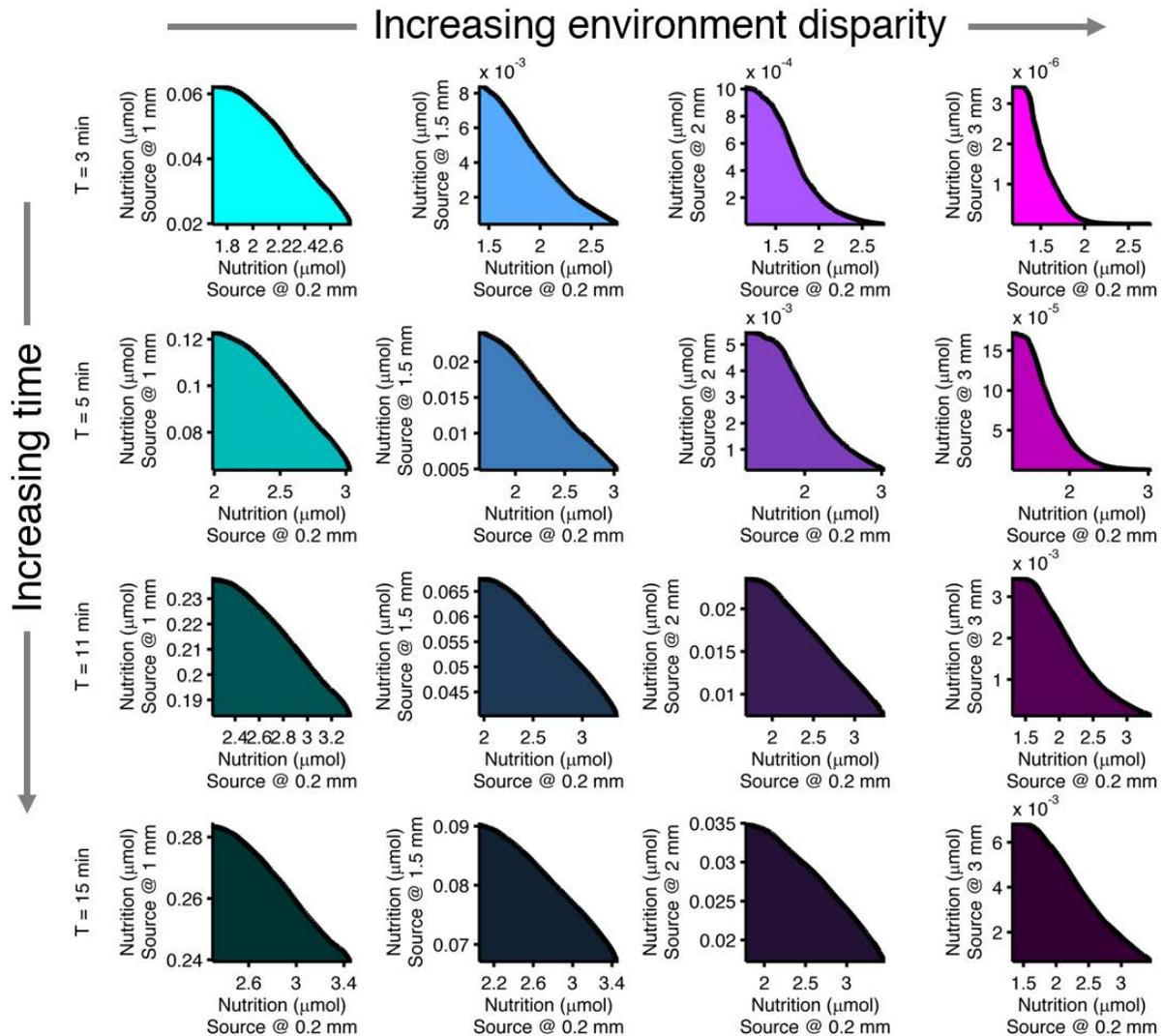

**Figure 2 – figure supplement 6.**
**Effect of time limits on near/far foraging trade-offs.**
Trade-offs in performance between foraging near and far sources are shown. From left to right (cyan to magenta), the far case is progressively more distant compared to the near case: 1mm, 1.5mm, 2mm, 3mm. From top to bottom (bright to dark colors), the time allotted is increasing: 3 min, 5min, 11min, 15min. Reduced time allotment makes the front (black line) more concave for the same pair of environments.

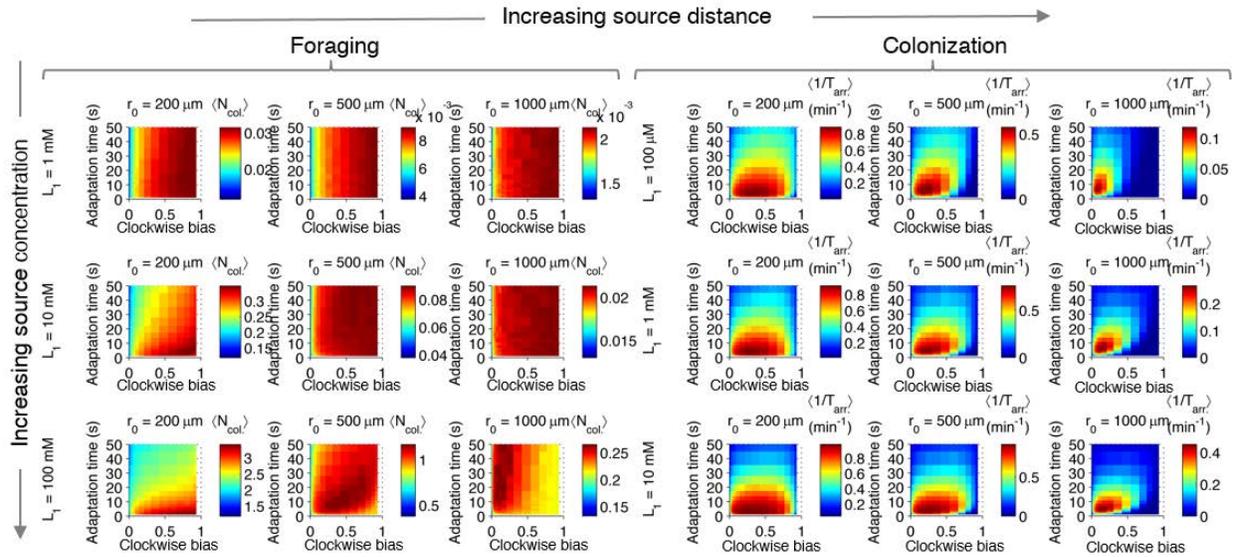

**Figure 2 – figure supplement 7.**
**Effect of source concentration on performance.**
Performance calculate and plotted as described in Figure 2 – figure supplement 3, but for different concentrations at the source. Left block ("Foraging"): foraging performance for increasing source distance (columns) and increasing source concentration (rows): $L_1$ = 1 mM, 10 mM, 100 mM. Right block ("Colonization") colonization performance for increasing source distance (columns) and increasing source concentration (rows): $L_1$ = 100 µM, 1 mM, 10 mM.

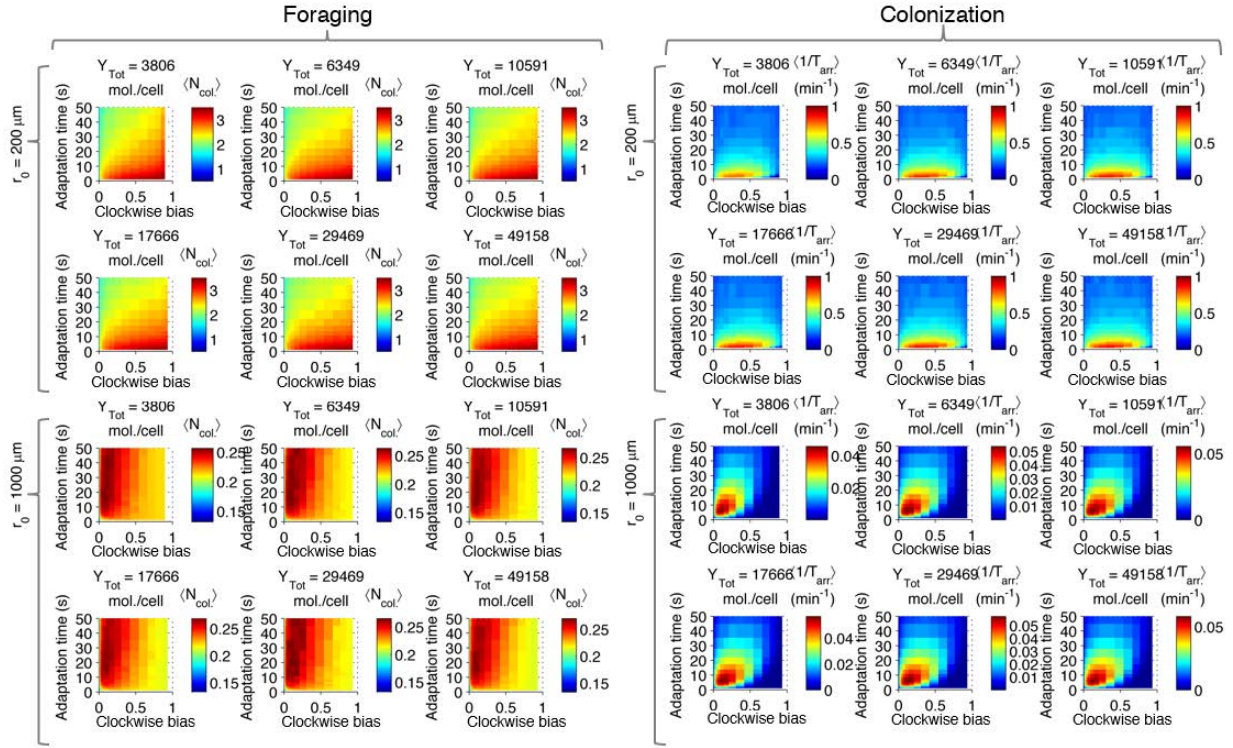

**Figure 2 – figure supplement 8.**
**Effect of CheY-P dynamic range on performance.**
Left block ("Foraging"): foraging performance for near (200 µm) and far (1000 µm) sources and increasing CheY-P dynamic range, which was changed through the total number of CheY molecules, $Y_{tot}$, as described in the SI. Right block ("Colonization") same as left block but for colonization.

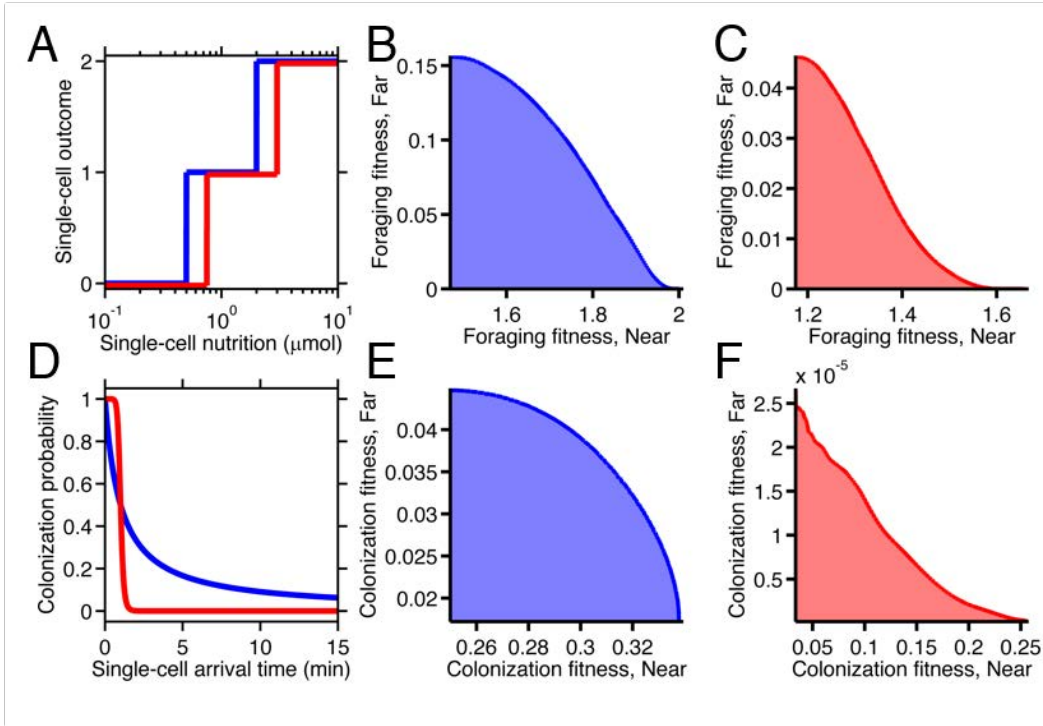

**Figure 5 – figure supplement 1.**
**Fitness trade-offs under alternate models of selection.**
**A**. Model of discrete physiological transitions applied to the chemotactic foraging challenge. Each individual replicate is given a number of progeny (0, 1, or 2) based on a two-step function of the nutrition they achieve from chemotaxis. For each phenotype, the foraging fitness is the average progeny across replicates. The effect of more (red) and less (blue) stringent nutrient requirements are compared. Survival requirement: 0.5 µmol (blue), 0.75 µmol (red), Division requirement: 2 µmol (blue), 3 µmol (red). **B–C**. Beginning with the foraging performance trade-off in Figure 4B, application of the survival model in A gives rise to either a weak (B) or strong (C) fitness trade-off, depending on where the thresholds and steepness are low (blue curve in A) or high (red curve in A). **D**. Probabilistic model of survival applied to the chemotactic colonization challenge. Each individual replicate survives has chance to survive depending on how soon it arrives. For each phenotype, the colonization fitness is the probability to colonize measured over all replicates. The effect of more (red) and less (blue) stringent survival functions are compared. Time threshold in both cases is 1 min with dependency 1 (blue) or 10 (red). **E–F**. Beginning with the arrival performance trade-off in Figure 4E, application of the selection model in C gives rise to either a weak (E) or strong (F) fitness trade-off.